%% file: Mn3X_main.tex
\documentclass[twocolumn,secnumarabic,superscriptaddress,amssymb, nobibnotes, aps, prb]{revtex4-2}

\usepackage{amsmath}
\usepackage{mathrsfs}  
\usepackage{mathtools} 
\usepackage{bbold}    
\usepackage{braket}    
\usepackage{gensymb}   

\usepackage{graphicx}      	
\usepackage{multirow}       
\usepackage{here} 			
\usepackage{longtable}

\usepackage{url}		
\usepackage[plainpages=false,pdfpagelabels=true, breaklinks]{hyperref}  
\hypersetup{colorlinks=false}

\usepackage{soul}                   
\usepackage[dvipsnames]{xcolor}		

\usepackage{siunitx}
\usepackage{xspace}

\input{new_commands}

\begin{document}

\title{Strong Magneto-Elastic Coupling in \mnx\ (X = Ge, Sn)}%

\author{Florian Theuss}%
\affiliation{Laboratory of Atomic and Solid State Physics, Cornell University, Ithaca, NY 14853, USA}
\author{Sayak Ghosh}
\affiliation{Laboratory of Atomic and Solid State Physics, Cornell University, Ithaca, NY 14853, USA}
\author{Taishi Chen}
\affiliation{The Institute for Solid State Physics, The University of Tokyo, Kashiwa, Chiba 277-8581, Japan}
\author{Oleg Tchernyshyov}
\affiliation{Institute for Quantum Matter and Department of Physics and Astronomy, Johns Hopkins University, Baltimore, MD 21218, USA}
\author{Satoru Nakatsuji}
\affiliation{Department of Physics, The University of Tokyo, Tokyo  113-0033, Japan}
\affiliation{The Institute for Solid State Physics, The University of Tokyo, Kashiwa, Chiba 277-8581, Japan}
\affiliation{Institute for Quantum Matter and Department of Physics and Astronomy, Johns Hopkins University, Baltimore, MD 21218, USA}
\affiliation{Trans-scale Quantum Science Institute, University of Tokyo, Tokyo 113-0033, Japan}
\author{B. J. Ramshaw}
\email{bradramshaw@cornell.edu}
\affiliation{Laboratory of Atomic and Solid State Physics, Cornell University, Ithaca, NY 14853, USA}

\date{\today}%

\begin{abstract}
We measure the full elastic tensors of \mnge and \mnsn as a function of temperature through their respective antiferromagnetic phase transitions. Large discontinuities in the bulk moduli at the N\'eel transitions indicate strong magnetoelastic coupling in both compounds. Strikingly, the discontinuities are nearly a factor of 10 larger in \mnge than in \mnsn. We use the magnitudes of the discontinuities to calculate the pressure derivatives of the N\'eel temperature, which are 39 K/GPa 14.3 K/GPa for \mnge and \mnsn, respectively. We measured in-plane shear modulus $c_{66}$, which couples strongly to the magnetic order,  in magnetic fields up to 18 T and found quantitatively similar behavior in both compounds. Recent measurements have demonstrated strong piezomagnetism in \mnsn: our results suggest that \mnge may be an even better candidate for this effect. 
\end{abstract}

\maketitle

\input{Introduction}

\input{methods}

\input{data}

\input{jumps_in_elastic_constants}

\input{pulse_echo}


\input{summary}


\section*{Acknowledgments}
The work at the Institute for Quantum Matter, an Energy Frontier Research Center, was funded by DOE, Office of Science, Basic Energy Sciences under Award \# DE-SC0019331. This work was partially supported by JST-Mirai Program (JPMJMI20A1), JST-CREST (JPMJCR18T3). This work made use of the Cornell Center for Materials Research (CCMR) Shared Facilities, which are supported through the NSF MRSEC Program (No. DMR-1719875).

\bibliography{literature}

\newpage
\onecolumngrid
\input{Mn3X_Supplementary}

\end{document}

%% file: new_commands.tex
\newcommand{\mnx}{Mn\textsubscript{3}X\xspace}
\newcommand{\mnge}{Mn\textsubscript{3}Ge\xspace}
\newcommand{\mnsn}{Mn\textsubscript{3}Sn\xspace}

\newcommand{\mnzerosn}{Mn\textsubscript{3.019}Sn\textsubscript{0.981}\xspace}

%% file: Introduction.tex
\section{Introduction}

Elastic strains offer a fast, local, and reversible way to manipulate the magnetic properties of solids. On a microscopic level, strains alter bond distances and the angles between magnetic ions, leading to changes in magnetic exchange coupling and magnetic anisotropy \cite{song_how_2018}. On a phenomenological level, these effects can lead to a strain dependence of the critical temperature and of the total magnetic moment. In the most extreme case, externally applied strains can break the crystal symmetry and drive magnetic phase transitions. The strain-dependence of the magnetization most commonly comes in the form of either magnetostriction, piezomagnetism, or flexomagnetism. All of these effects find useful applications in the recently-emerging field of straintronics \cite{miao_straintronics_2021, bukharaev_straintronics_2018}. This necessitates the search for materials with large magnetoelastic coupling.

In this regard, the noncollinear antiferromagnets \mnx (X = Ge, Sn) are promising candidates. 120$\degree$ triangular magnetic order forms in these compounds well above room temperature. This magnetic order is the source of several anomalous transport properties including giant anomalous Hall, Nernst, and thermal Hall effects \cite{Kiyohara2016,Nayak2016,Ikhlas2017,Guo2017,Wuttke2019a,Nakatsuji2015,Sung2018,Li2017}. These quantities were recently shown to be strongly strain dependent. For example, \citet{Reis2020} demonstrated the ability to change the sign of the Hall angle in \mnge by applying hydrostatic pressure, and \citet{Ikhlas2022} switched the direction of the Hall effect in \mnsn by applying uniaxial strain. Additional evidence for large magnetoelastic coupling has been found in neutron diffraction studies \cite{Sukhanov2019}, as well as in spontaneous magnetostriction at $T_N$ \cite{Sukhanov2018} in \mnge. Most recently, \mnsn was found to exhibit an extraordinarily large piezomagnetic effect \cite{Ikhlas2022}.These findings reveal an intimate connection between magnetism, anomalous transport properties, and elastic strain in \mnx, making it a prime candidate for applications in straintronics. 

While many anomalous transport coefficients have been documented in \mnx, the fundamental quantity relating stress and strain---the elastic tensor---has not been measured. From a practical standpoint, the elastic moduli are needed to convert stresses---the quantity typically known in an experiment---to strains. From a fundamental standpoint, elastic moduli are a powerful thermodynamic probe into the symmetry breaking at the magnetic phase transition. 


We directly measure the full elastic tensors of \mnge and \mnsn through the respective phase transitions. We study the elastic moduli using resonant ultrasound spectroscopy (RUS) and pulse-echo ultrasound. We find large discontinuities at $T_N$ in the compressional elastic moduli and, using Ehrenfest relations, relate them to large derivatives of the N\'eel temperature with respect to hydrostatic pressure. We calculate $dT_N/dP$ to be roughly 39 K/GPa in \mnge and 14.3 K/GPa in \mnsn---some of the largest values ever reported for itinerant antiferromagnets. We measure $c_{66}$---corresponding to the strain that switches the sign of the anomalous Hall coefficient \cite{Ikhlas2022}---in magnetic fields up to 18 tesla. We find that, while the elastic moduli of \mnge and \mnsn exhibit large quantitative differences in zero field, their magnetic field dependencies are quite similar.

This paper is structured as follows: in \autoref{section: methods} we describe our pulse-echo and RUS measurements and how our data is analyzed. In \autoref{section: data} we describe the measured elastic moduli and their temperature dependencies. We analyze the temperature dependencies quantitatively in \autoref{section: analysis}. We analyze the magnetic field dependence of $c_{66}$ in \autoref{section: pulse_echo}. Finally, we summarize our conclusions in \autoref{section:summary}.

%% file: methods.tex
\section{Methods}
\label{section: methods}

\begin{figure}
	\centering
	\includegraphics[width=1\linewidth]{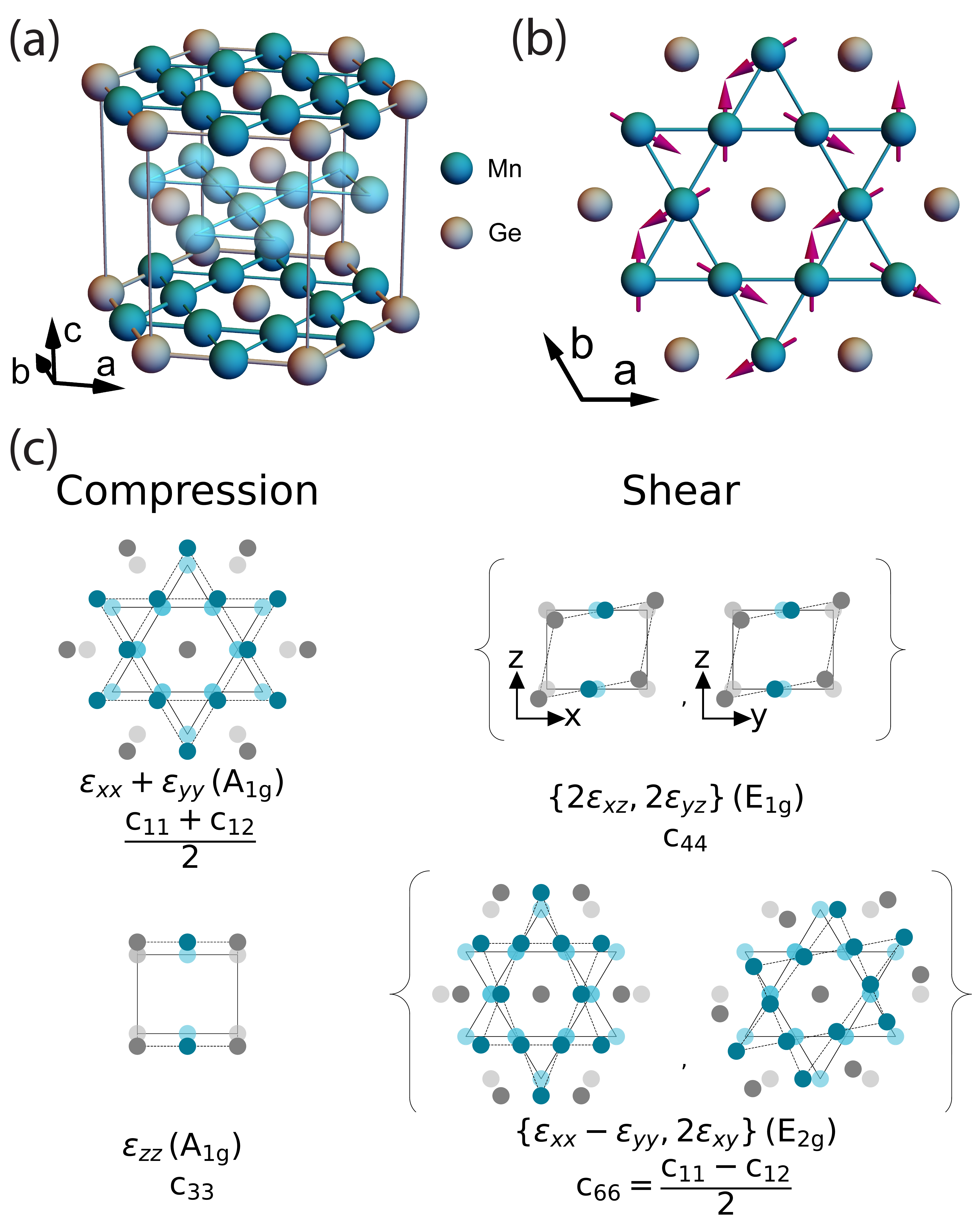}
	\caption{\textbf{Crystal structure and irreducible strains of \mnx.} a) \mnx~ crystal structure. A hexagonal unit cell consists of AB-stacked Kagome planes of Mn atoms. Different shades of green indicate A and B planes, respectively. b) View of one Mn Kagome layer, with purple arrows that illustrate one possible ordered-state spin configuration. c) Visualization of the irreducible representations of strain. The definition of irreducible strains are given in terms of $\varepsilon_{ij}$, alongside the symmetry representations and the corresponding elastic moduli.}
	\label{fig:irreducible strain visualization}
\end{figure}

\mnx (X = Ge, Sn) crystallizes with a hexagonal unit cell (point group $D_{6h}$ \autoref{fig:irreducible strain visualization} a), with lattice parameters $a = \SI{5.3}{\angstrom}$ and $c = \SI{4.3}{\angstrom}$ for \mnge \cite{qian_exchange_2014}, and $a = \SI{5.7}{\angstrom}$ and $c = \SI{4.5}{\angstrom}$ for \mnsn \cite{markou_noncollinear_2018}. Mn atoms form a Kagome lattice in the a-b plane, and local moments on the Mn sites order in a chiral antiferromagnetic stucture (\autoref{fig:irreducible strain visualization} b) \cite{Nagamiya1982,Chen2020}, with a small in-plane magnetic moment due to spin canting \cite{Tomiyoshi1983}. Neutron diffraction studies find a magnetic order parameter of the $E_{1g}$ representation in the $D_{6h}$ point group \cite{Soh2020a,Chen2020}. The N\'eel temperature ($T_N$) for \mnge is 370~K. For \mnsn, $T_N$ depends strongly on the exact stochiometry: here we investigated \mnzerosn with a critical temperature of 415~K. This composition of \mnsn features an additional phase transition to spiral spin order below about 270~K \cite{kren_study_1975,cable_neutron_1993}. For simplicity, we will refer to \mnzerosn as \mnsn for the remainder of this paper.

The strain tensor in $D_{6h}$ consists of four independent elements. Linear combinations of these form irreducible representations (irreps, \autoref{fig:irreducible strain visualization}c). The irreps are divided into two one-component compressional strains that transform as the $A_{1g}$ irrep, and two two-component shear strains transforming as the $E_{1g}$ (out-of-plane shear) and $E_{2g}$ (in-plane shear) irreps. The elastic moduli corresponding to each irrep are defined according to $c_\Gamma = \partial^2 \mathcal{F} / \partial \varepsilon_\Gamma^2$, where $\mathcal{F}$ is the free energy and $\Gamma$ labels the irreps. The resulting elastic moduli are $c_{A1g,1}=\left(c_{11}+c_{12}\right)/2$, $c_{A1g,2} = c_{33}$, $c_{E1g}=c_{44}$, and $c_{E2g} = c_{66} = \left(c_{11}-c_{12}\right)/2$. An additional, fifth elastic modulus, $c_{A1g,3}=c_{13}$, couples the in-plane and out-of-plane compressional strains. \autoref{fig:irreducible strain visualization}c illustrates these irreducible strains, provides their definitions in terms of the strains $\varepsilon_{ij}$, and gives the corresponding elastic moduli.

We measured the temperature dependence of the the full elastic tensor using resonant ultrasound spectroscopy (RUS). In RUS, a sample is placed on its corners in weak mechanical contact between two piezoelectric transducers to provide nearly-free elastic boundary conditions. One transducer is driven at a fixed frequency and the resulting charge generated at the other transducer is detected using a custom-built charge amplifier and digital lockin (the amplifier and lockin are described in \citet{balakirev2019resonant}). By sweeping the drive frequency in the range of 0.1 to 5 MHz, we can measure the lowest mechanical resonance frequencies of a three-dimensional solid. From these resonance frequencies, we then determine the elastic moduli by inverse-solving the elastic wave equation (see \cite{Ramshaw2015a,Ghosh2020b} for details on technique and data analysis). In contrast to the conventional pulse-echo ultrasound technique, where only one elastic modulus is measured at a time, RUS allows the extraction of the temperature dependence of the full elastic tensor with one experiment. 

To access the relatively high N\'eel temperatures of \mnx, we built an RUS apparatus inside an insulated box on a hotplate (see SI for pictures of the measurement setup). The temperature was monitored with a Lakeshore PT100 platinum resistance thermometer and recorded with a Cryocon Model 22C temperature controller. For the fits of the elastic tensor, we used the lowest resonance frequencies up to 4~MHz in both compounds. This included 84 resonances for \mnge and 68 for \mnsn. Our fits converged with root mean square errors of 0.18~\% (387~K) and 0.42~\% (300~K) for \mnge, and 0.23~\% (438~K) and 0.48~\% (300~K) for \mnsn. More details, including a full list of experimental and calculated resonances, can be found in the SI.
 
The requirement of weak mechanical contact between transducers and the sample makes it difficult to reliably perform RUS in magnetic fields. To measure the $c_{66}$ elastic modulus as a function of magnetic field, we employed the pulse-echo ultrasound technique \cite{luthi_physical_2005}. Ultrasound waves were generated by 41$\degree$ X-cut LiNbO\textsubscript{3} shear transducers with a fundamental frequency of 40 MHz, purchased from Boston Piezo-Optics Inc. The transducers were driven at 199 MHz for \mnge and at 175 MHz for \mnsn. The transducers were glued to polished surfaces of the sample, perpendicular to the c-axis, using Angstr\"omBond AB9110LV from Fiber Optic Center Inc. Short (80 ns) bursts of ultrasound were generated from the transducer using a Tektronix TSG 4106A RF generator and amplified with a Mini-Circuits ZHL-42W+ power amplifier. The ultrasonic echoes are detected using the same transducer, amplified with a Mini-Circuits ZX60-3018G-S+ amplifier and captured on a Tektronix MSO64 oscilloscope. A software lockin is used to track phase changes in the echoes as a function of temperature and magnetic field, allowing relative changes in the sound velocity, $\Delta v/v$, to be measured with an accuracy of  better than $10^{-6}$. In this configuration, we measure changes in $v_{66}$, which are converted to the associated elastic modulus change by $\Delta c_{66}/c_{66} = 2 \Delta v_{66} / v_{66}$.
 
The pulse-echo measurements were performed with a custom high-temperature probe in an Oxford Instruments variable temperature insert (VTI) in an Oxford Instruments 20 Tesla superconducting magnet system. The sample space of the VTI was pumped continuously throughout the experiment to ensure high vacuum. We performed these measurements with an in-plane magnetic field applied parallel to the polarization vector of the sound wave (and perpendicular to the direction of sound propagation).

%% file: data.tex
\section{Data}
\label{section: data}

\autoref{table:high_T_elastic_constants} lists the elastic moduli of \mnge and \mnsn at room temperature and at high temperatures---above their respective antiferromagnetic phase transitions---as well as their bulk moduli and Poisson's ratios. In their respective paramagnetic states, the compresional elastic moduli, $\left( c_{11}+c_{12} \right) /2$ and $c_{33}$, are 13 \% and 28 \% larger in \mnge than in \mnsn. This implies tighter bonding in \mnge, which is also consistent with its smaller unit cell. The value of the in-plane Poisson ratio, $\nu_{xy}$, is consistent with what is found in most conventional metals \cite{Davis1998}. $\nu_{zx}$ on the other hand, is anomalously small, even compared to other layered materials like Sr\textsubscript{2}RuO\textsubscript{4} ($\nu_{zx}=0.16$ \cite{Ghosh2020b}), URu\textsubscript{2}Si\textsubscript{2} ($\nu_{zx}=0.20$ \cite{Ghosh2020}), CeIrIn\textsubscript{5} ($\nu_{zx}=0.32$ \cite{bachmann_spatial_2019}), and La\textsubscript{2}CuO\textsubscript{4} ($\nu_{zx}=0.21$ \cite{migliori_elastic_1990}), implying extremely weak elastic coupling between different planes in the hexagonal crystal structure of \mnx.


\begin{table*}[]
	\begin{tabular}{|cw{c}{1.9cm}|w{c}{1.4cm}w{l}{.2cm}w{l}{1.2cm}w{l}{.1cm}w{l}{1.3cm}w{l}{.2cm}w{l}{1.2cm}w{c}{1.4cm}|w{c}{2.1cm}|w{c}{1.4cm}w{c}{1.4cm}|}
		\hline
		&  & \multicolumn{8}{c|}{} &  & \multicolumn{2}{c|}{} \\[-1em]
		Compound                   & Temperature & \multicolumn{8}{c|}{Elastic Moduli (GPa)} & Bulk Modulus & \multicolumn{2}{c|}{Poisson's Ratios} \\
		                           & (K)		 & $\frac{c_{11}+c_{12}}{2}$ & \multicolumn{2}{c}{$c_{13}$} & \multicolumn{2}{c}{$c_{33}$} & \multicolumn{2}{c}{$c_{44}$} & $\frac{c_{11}-c_{12}}{2}$ & $B$ (GPa) & $\nu_{xy}$ & $\nu_{zx}$\\
		&  & \multicolumn{8}{c|}{} &  & \multicolumn{2}{c|}{} \\[-1em]
		\hline
		\hline
		&  & \multicolumn{8}{c|}{} &  & \multicolumn{2}{c|}{} \\[-1em]
		\multirow{2}{*}{\mnge}     & 300         & 87.0(5) && 12.5(15) && 201.5(16) && 48.4(1) & 43.0(5) & 65.9(7) & 0.334(6) & 0.041(5)\\ 
								   & 387         & 90.4(2) && 14.6(6)  && 194.6(5)  && 45.09(5) & 48.1(2) & 67.9(3) & 0.300(2) & 0.053(2)\\
		&  & \multicolumn{8}{c|}{} &  & \multicolumn{2}{c|}{} \\[-1em]
		\hline
		&  & \multicolumn{8}{c|}{} &  & \multicolumn{2}{c|}{} \\[-1em]
		\multirow{2}{*}{\mnsn}     & 300         & 85.8(5) && 18.1(14) && 165.3(11) && 52.0(2) & 50.8(5) & 64.5(7) & 0.246(7) & 0.083(7)\\ 
								   & 438         & 79.7(2) && 17.0(6)  && 151.3(5)  && 48.11(8) & 51.2(2) & 59.7(3) & 0.206(3) & 0.089(3)\\
		\hline
	\end{tabular}
	\caption{\textbf{Elastic properties of \mnx}. All quantities are reported at 387 K for \mnge and at 438 K for \mnsn, where each compound is in the paramagnetic state, as well as at room temperature. Elastic moduli, as well as the bulk modulus, are given in GPa. The Poisson's ratios $\nu_{xy}$ and $\nu_{zx}$ are also given. The definitions of these Poisson's ratios in terms of elastic moduli are given in the SI. In a hexagonal crystal $c_{66} = \frac{c_{11}-c_{12}}{2}$. We show full temperature dependences of the Poisson's ratios and the bulk modulus in the SI.}
	\label{table:high_T_elastic_constants}
\end{table*}

\begin{figure*}
	\centering
	\includegraphics[width=1\linewidth]{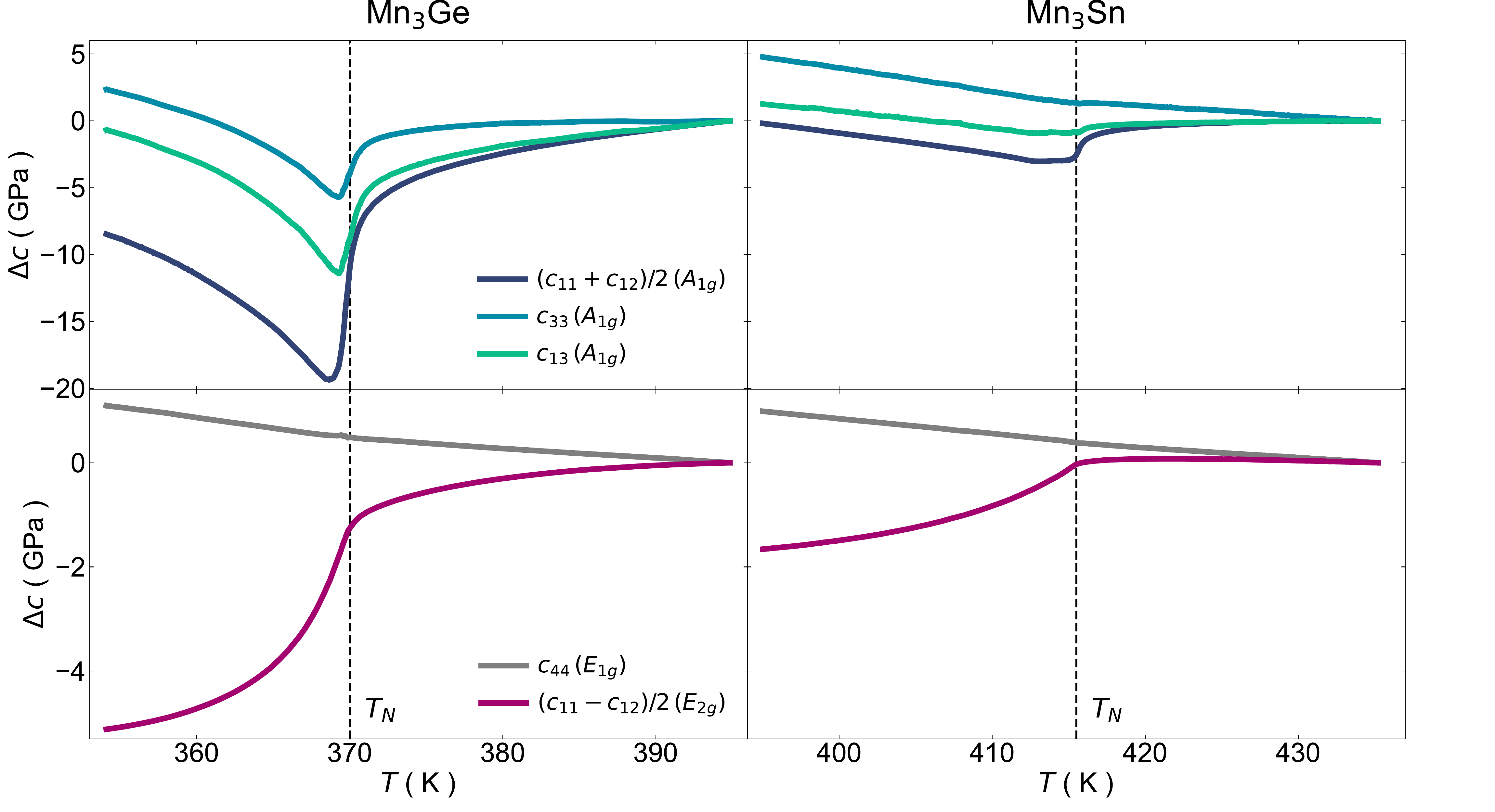}
	\caption{\textbf{Change in elastic moduli as a function of temperature for \mnge (left) and \mnsn (right).} Upper and lower panels show changes compressional and shear elastic moduli, respectively. The change is defined as $\Delta {\rm c}(T) = {\rm c}(T)-{\rm c}(387 K)$ for \mnge and $\Delta {\rm c}(T) = {\rm c}(T)-{\rm c}(438 K)$ for \mnsn. The N\'eel temperatures are indicated by vertical dashed lines.}
	\label{fig:elastic constants}
\end{figure*}

To investigate the coupling between magnetism and elasticity in \mnx, we first measured the elastic moduli as a function of temperature through their respective N\'eel temperatures $T_N$ (see \autoref{fig:elastic constants}).

We first discuss the temperature dependence of the compressional elastic moduli (upper panels of \autoref{fig:elastic constants}). Starting well above $T_N$, all three compressional moduli in \mnge decrease monotonously upon cooling towards the phase transition. This anomalous softening is in contrast to the conventional stiffening of elastic moduli when the temperature is lowered \cite{Varshni1970}, and implies sizable antiferromagnetic fluctuations well above $T_N$. These fluctuations suggest a non-mean-field phase transition in \mnge, as there are no precursor fluctuations above $T_N$ in mean field theory. The softening of the compressional moduli above $T_N$ is followed by a step-like feature at the phase transition. 

Qualitatively similar behavior is seen in \mnsn, but with quantitative differences. In \mnsn, $c_{13}$ and $\left(c_{11}+c_{12} \right)/2$ are almost temperature independent well above $T_N$, and $c_{33}$ increases upon cooling. All compressional elastic moduli eventually soften above $T_N$, but much more weakly than in \mnge. Additionally, the absolute sizes of the steps at $T_N$ are nearly a factor of 10 smaller in \mnsn than in \mnge. Both the smaller precursor softening and the smaller steps at $T_N$ suggest that the coupling between magnetism and the lattice is significantly stronger in \mnge than in \mnsn.

We now turn to the shear moduli (lower panels in \autoref{fig:elastic constants}). The behavior of $c_{44}$ is relatively conventional, with no precursor softening and only a change in slope at $T_N$. $c_{66}$, on the other hand, softens towards the N\'eel temperature upon cooling, similar to the compressional modes. The much stronger signature in \mnge than in \mnsn again indicates stronger magnetoelastic coupling in the former compound. While a step in $c_{66}$ at $T_N$ is allowed by symmetry for the chiral order in \mnx \cite{dasgupta_theory_2020,Ikhlas2022}, no feature that is comparable in width to the steps in the compressional moduli is seen in $c_{66}$ (see SI for a derivation of which moduli can show discontinuous jumps at the phase transition). Note that the $E_{2g}$ strain associated with $c_{66}$ is the same strain that is responsible for the piezomagnetic effect and the switching of the anomalous Hall effect. The precursor softening in this channel again indicates the non-mean-field nature of the magnetic phase transition in \mnx, and will be the subject of a future study.

%% file: jumps_in_elastic_constants.tex
\section{Discontinuities in Compressional Elastic Moduli at $T_N$}
\label{section: analysis}

The discontinuities in the compressional moduli at $T_N$ are indicative of a second-order phase transition and are reminiscent of a heat capacity anomaly \citet{Rehwald1973}. Indeed, Ehrenfest relations require that the changes in the heat capacity and the compressional moduli across $T_N$ are proportional to each other, and the coefficient of proportionality is the square of the derivative of $T_N$ with respect to hydrostatic pressure $P_{\rm hydro}$. Using the measured heat capacity and our measurements of the compressional moduli, we can calculate $dT_N/dP_{\rm hydro}$.

The Ehrenfest relation between the bulk modulus and heat capacity discontinuities is \cite{Ghosh2020b}
\begin{equation}
\left( \frac{dT_N}{dP_{\rm hydro}} \right)^2 = - \frac{\Delta B}{B^2} \left( \frac{\Delta C}{T_N} \right)^{-1},
\label{eq:Ehrenfest relation}
\end{equation}
where $\Delta B$ and $\Delta C$ are the sizes of the discontinuities in the bulk modulus and specific heat, respectively, and $B$ is the absolute bulk modulus at $T_N$. 

\begin{figure}
	\centering
	\includegraphics[width=1\linewidth]{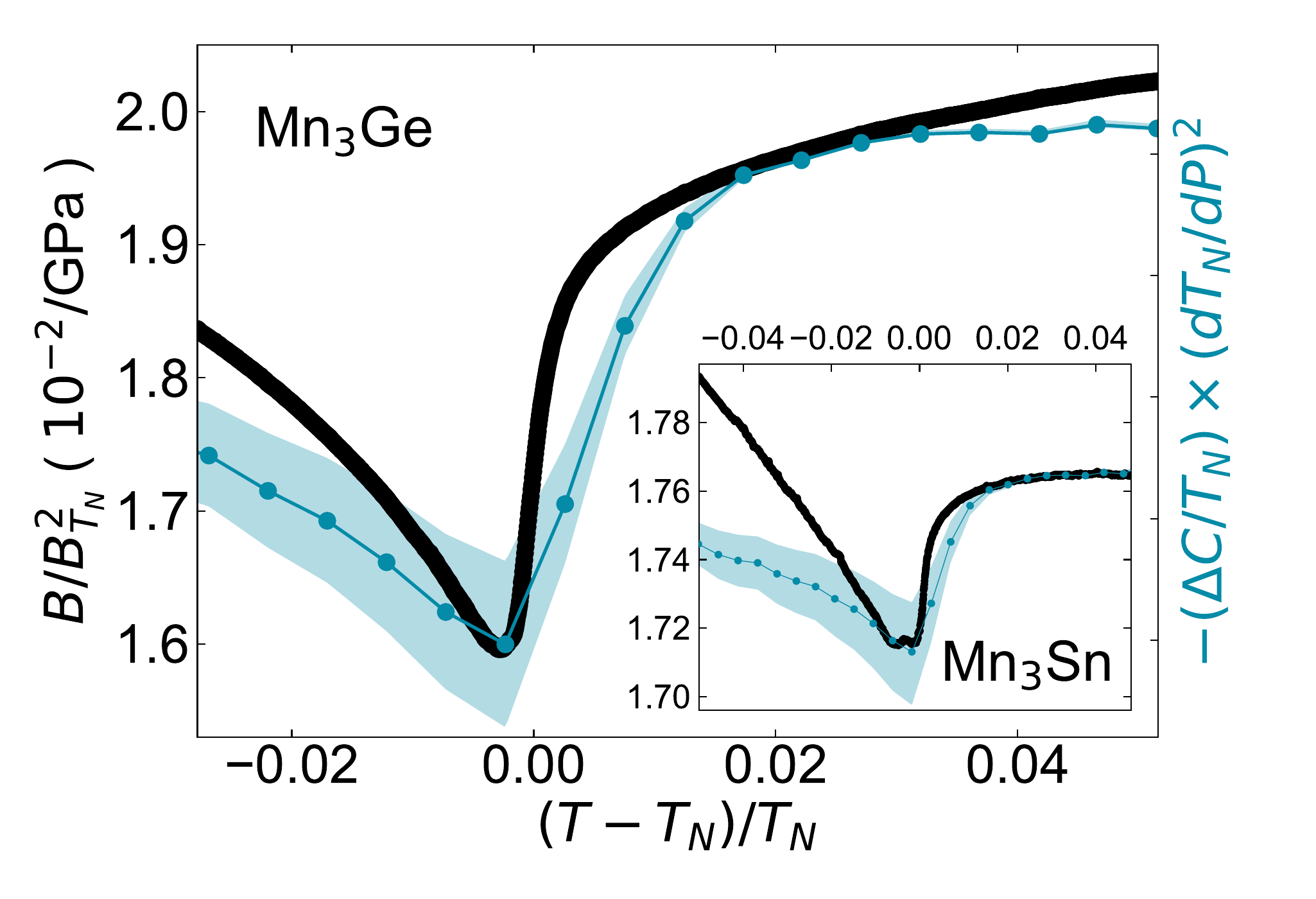}
		\caption{\textbf{Ehrenfest scaling for the bulk moduli and specific heat of \mnx}. Blue points are the specific heat of \mnge taken from \cite{Chen2020} divided by the N\'eel temperature $T_N$ and scaled by a factor with units of $\rm (kelvin/GPa)^2$. The bulk modulus of \mnge, divided by $B_{T_N}^2$---the square of the value of the bulk modulus at $T_N$---is shown as black points in the main panel. Both data sets are given in the same units, i.e. 1/GPa. The scaling factor used here corresponds to a value of $dT_N/dP = 39\,\mathrm{K/GPa}$, and the shaded region corresponds to deviations of $\pm3\,\mathrm{K/GPa}$. In the inset, this analysis is repeated for the bulk modulus of \mnsn. It reflects a value of  $dT_N/dP \approx \left( 14.3 \pm 2 \right)\,\mathrm{K/GPa}$.}
	\label{fig:Ehrenfest relation}
\end{figure}

To extract the derivative of the N\'eel temperature with hydrostatic pressure from our data, we plot the bulk modulus $B / B \left( T_N\right)^2$ on the same scale as the specific heat scaled by $dT_N / dP_{hydro}$, i.e. $- \Delta C/T_N   \left( dT_N / dP_{hydro} \right)^2$ (see \autoref{fig:Ehrenfest relation} and footnote \footnote{For mean-field like transitions, the discontinuities in thermodynamic coefficients are easily defined as the difference in the coefficient immediately above and below the transition. In the case of \mnx, the transition does not appear mean-field like. We therefore use this scaling procedure to avoid ambiguity in the definition of the discontinuity, as the non-mean field ``rounding'' is approximately the same in the specific heat and the bulk modulus.}). This analysis for \mnge, along with the specific heat data for \mnge from \citet{Chen2020}, is shown in the main panel of \autoref{fig:Ehrenfest relation}. We extract a derivative of $T_N$ with respect to pressure of $dT_N / dP_{hydro} =39 \pm 3$ K/GPa.

Specific heat data are not available for \mnsn through its high temperature phase transition. However, using the specific heat data for \mnge, we estimate $dT_N/dP_{\rm hydro} \approx \left( 14.3 \pm 2 \right)\,\mathrm{K/GPa}$ for \mnsn (see inset of \autoref{fig:Ehrenfest relation}). This value is about a factor of three smaller than for \mnge. It is possible that the true heat capacity of \mnsn is a factor of 9 larger than in \mnge. Either way---whether it is due to a factor of 9 difference in heat capacity or a factor of 3 difference in $dT_N/dP_{\rm hydro}$---this observation is puzzling given that the two compounds share similar values of $T_N$, the same room-temperature magnetic structure, and the same crystal structure with only marginally different unit cell parameters.

\autoref{table:compare_jump_magnitude_between_other_antiferromagnets} compares the size of $d T_N/ dP_{hydro}$ between several metallic antiferromagnets. \mnge and \mnsn stand out with some of the largest pressure derivatives of their respective N\'eel temperatures. Only the alloy Mn\textsubscript{3}Pt and elemental chromium have transition temperatures more sensitive to pressure than \mnge. Note that these compounds and \mnx are also the only materials with transitions above room temperature. These features, as well as their metallic conductivity, make \mnge and \mnsn two of only a few materials exceptionally well suited for applications in straintronics.
\begin{table}[]
	\begin{tabular}{w{c}{1.8cm}w{c}{1.8cm}w{c}{1cm}w{c}{1.8cm}}
		\hline
		\\[-1em]
		\multirow{3}{*}{Compound}	& $\frac{dT_N}{dP_{hydro}}$ & $T_N$ 	&  \multirow{3}{*}{Reference} \\
		\\[-1em]
		                            & (K/GPa)                 & (K)     & \\
		\\[-1em]
		\hline
		\hline
		\\[-1em]
		\mnge	 					& 39									& 370			& This work\\
		\mnsn 	 					& 14.3									& 415			& This work\\
		\\[-1em]
		\hline
		\\[-1em]
		Mn\textsubscript{3}Pt		& 70									& 475			& \cite{yasui_pressure_1987}\\
		Cr       					& 51									& 312			& \cite{mcwhan_pressure_1967}\\
		$\alpha$-Mn 				& 17									& 95			& \cite{mori_effect_1972}\\
		UN							& 9.3									& 53			& \cite{nakashima_high-pressure_2003}\\
		CuMnSb						& 4.7									& 50			& \cite{malavi_high-pressure_2018}\\
		MnPd\textsubscript{3}		& 2.0									& 195			& \cite{yasui_pressure_1988}\\
		UPtGa\textsubscript{5}	 	& 1.5									& 26			& \cite{nakashima_high-pressure_2003}\\
		CrB\textsubscript{2}		& 1.0									& 87			& \cite{grechnev_effect_2009}\\
		TiAu					  	& 0.6									& 33			& \cite{wolowiec_pressure_2017}\\
		UIrGe					  	& 0.11									& 16.5			& \cite{pospisil_effect_2017}\\
		\hline
	\end{tabular}
	\caption{ The derivative of the N\'eel temperature with respect to hydrostatic pressure for selected metallic antiferromagnets.}
	\label{table:compare_jump_magnitude_between_other_antiferromagnets}
\end{table}

%% file: pulse_echo.tex
\section{$c_{66}$ in Magnetic Field}
\label{section: pulse_echo}


The in-plane shear strain, $\boldsymbol{\varepsilon}_{E2g} = \{ \varepsilon_{xx}-\varepsilon_{yy}, 2 \varepsilon_{xy} \}$, plays a special role in the coupling between magnetism and strain in \mnx. Unlike most shear strains in magnetic systems, $\boldsymbol{\varepsilon}_{E2g}$ can couple to the magnetic order parameter $\boldsymbol{\eta} = \{ \eta_x, \eta_y \}$ as $((\varepsilon_{xx}-\varepsilon_{yy})(\eta_x^2-\eta_y^2) + 4 \varepsilon_{xy}\eta_x \eta_y)$ within a Landau free energy. This type of coupling---linear in shear strain and quadratic in order parameter---can reorient the magnetic moments on the Kagome lattice and align domains \cite{dasgupta_theory_2020,Ikhlas2022}. \citet{Ikhlas2022} used $\boldsymbol{\varepsilon}_{E2g}$ strain to change the sign of the Hall coefficient and to find a large piezomagnetic effect in \mnsn. This motivates a measurement of the associated elastic modulus, $c_{66}$, in external magnetic fields.

\begin{figure}[H]
	\centering
	\includegraphics[width=1\linewidth]{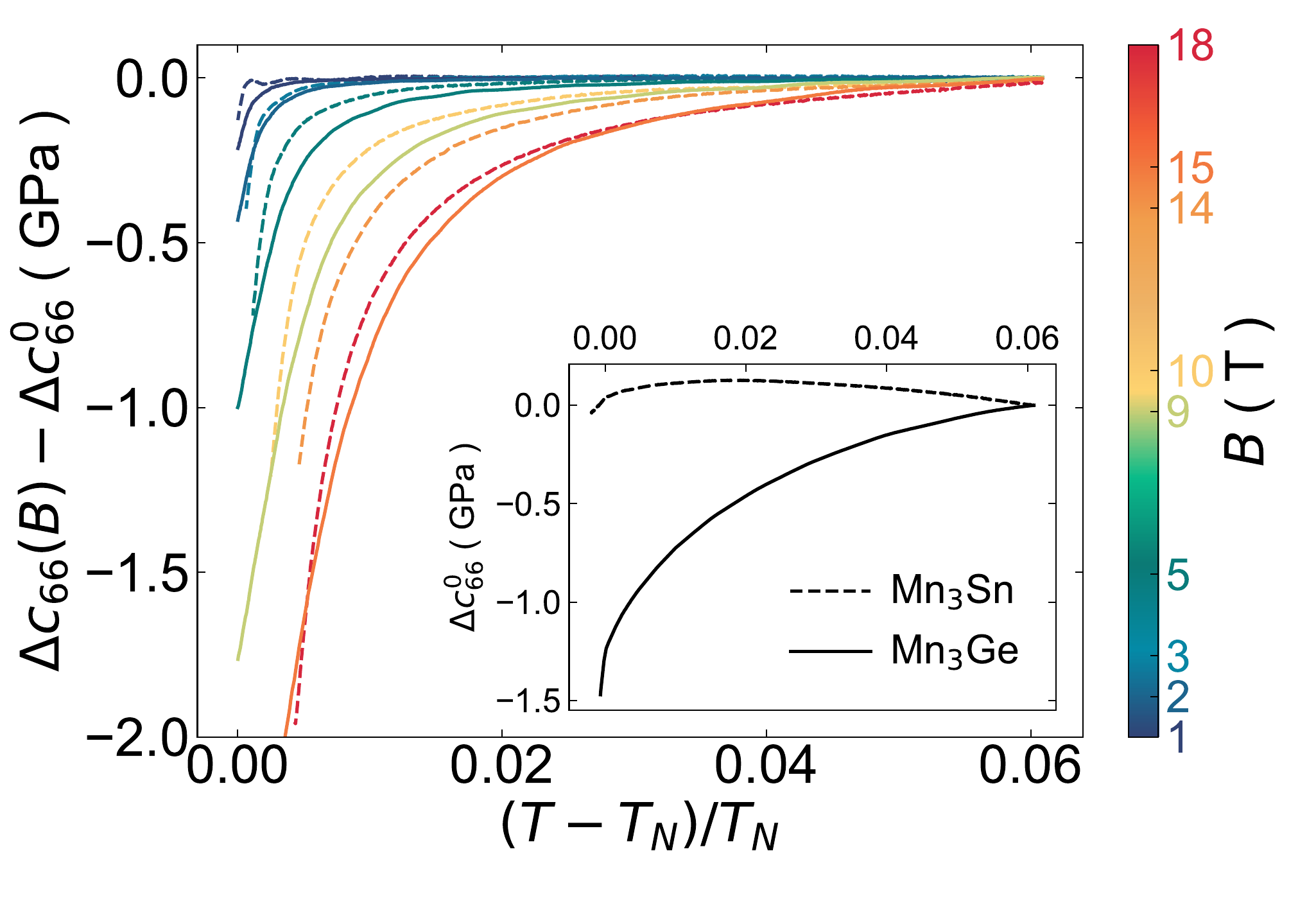}
	\caption{\textbf{The field dependence of $c_{66}$ in \mnx.} The changes in $c_{66}$ for \mnge (solid lines) and \mnsn (dashed lines) at different fields with respect to the zero-field elastic moduli are shown as a function of the reduced temperature. The data were taken at 1, 2, 5, 9, and 15~T for \mnge and at 1, 2.7, 5, 10, 14, and 18~T for \mnsn.
	The inset shows the zero-field data for both compounds.}
	\label{fig:pulse echo data}
\end{figure}

The inset to \autoref{fig:pulse echo data} shows the change in $c_{66}$ as a function of temperature in zero magnetic field for \mnge and \mnsn, measured with pulse-echo ultrasound. The main panel of \autoref{fig:pulse echo data} shows this temperature dependence at different magnetic fields with the zero-field data subtracted from each curve. The data are shown as a function of the reduced temperature $(T-T_N)/T_N$ above their respective phase transitions. The data end at (or just before) $T_N$ because the ultrasonic attenuation becomes too large to resolve a clear signal in the ordered phase.

As noted earlier, the temperature dependence of $c_{66}$ in zero field shows much stronger precursor fluctuations in \mnge than in \mnsn. However, once we account for this difference in the zero-field temperature dependence, the change with magnetic field is quite similar for the two compounds. With increasing magnetic field, the softening towards $T_N$ becomes more pronounced. This behavior is reminiscent of ferromagnetic transitions and is indicative of the trilinear coupling allowed by symmetry between shear strain, magnetic order parameter, and the external field in \mnx (see SI for a description of this coupling).

%% file: summary.tex
\section{Discussion}
\label{section:summary}

In summary, we used resonant ultrasound spectroscopy and pulse-echo ultrasound to measure the elastic moduli of \mnge and \mnsn. In addition to the full elastic tensor, we also provide the bulk moduli and Poisson's ratios. We find an anomalously small out-of-plane Poisson's ratio, $\nu_{zx}$, in both materials, implying weak elastic coupling between different layers of the hexagonal crystal structure. By scaling the bulk modulus anomalies to match the heat capacity anomaly at $T_N$, we extract large derivatives of the N\'eel temperatures with respect to hydrostatic pressure: $\left(39 \pm 3 \right)$~K/GPa and $\left(14.3 \pm 2.0 \right)$~K/GPa in \mnge and \mnsn, respectively. Finally, although the zero-field magneto-elastic coupling appears to be much larger in \mnge than in \mnsn, we find that the field dependence of the in-plane shear modulus---associated with the strain that couples strongly to the magnetism in \mnx---is similar in the two compounds.

The \mnx family hold promise for straintronic applications because it combines metallic conductivity, robust room-temperature magnetism, a large anomalous Hall effect whose sign can be switched with strain, and strong piezomagnetism. The latter two properties---piezomagnetism and strain dependence of anomalous transport properties \cite{Ikhlas2022}---have only been performed on \mnsn. Our measurements suggest that these effect may be even more dramatic in \mnge.
%



%% file: Mn3X_Supplementary.tex
\section{Supplementary Material}
\subsection{Resonant Ultrasound Spectroscopy (RUS)}
\begin{figure}
	\centering
	\includegraphics[width=\linewidth]{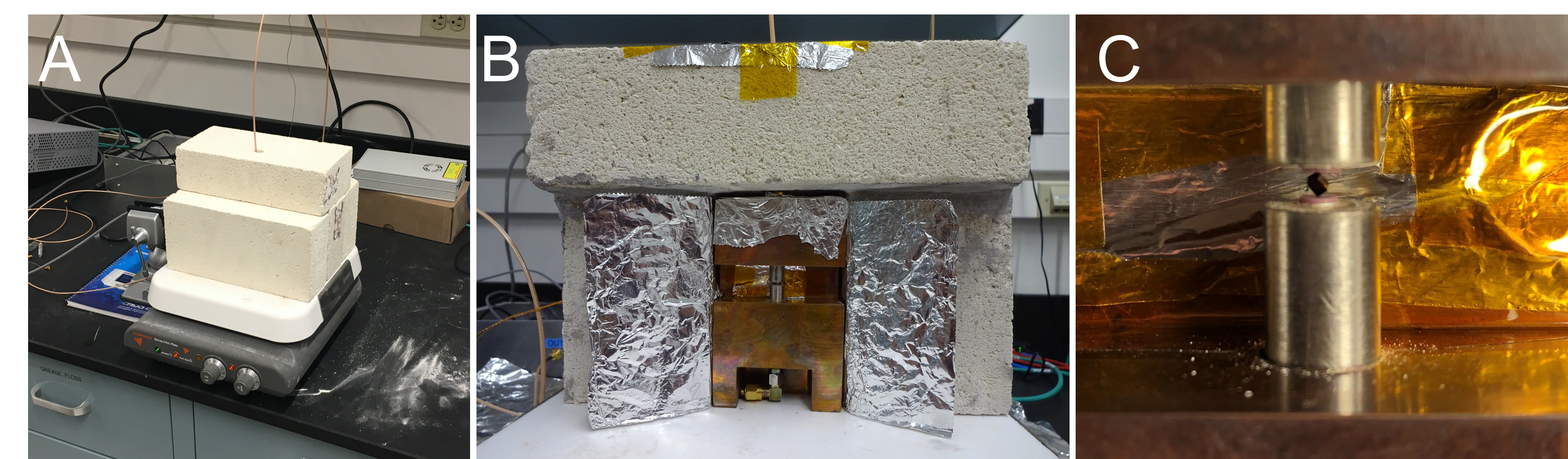}
	\caption{Pictures of the custom-built high temperature RUS setup. Panel A) shows the entire setup covered by insulating firebricks. In panel B), one of the firebricks has been removed to give a better view of the copper mount. Panel C) shows a sample being mounted on its corners between two piezoelectric transducers.}
	\label{fig: setup}
\end{figure}

To access temperatures above the N\'eel temperatures of \mnx, our RUS experiments were performed in a custom-built experimental setup (see \autoref{fig: setup}). It consists of a large copper mount placed on a hotplate and insulated with firebricks. Ultrasound was created by two compressional-mode lithium niobate transducers glued to stainless steel rods with ceramic epoxy. These transducer rods were placed in the copper mount such such that free vertical motion was allowed for the top transducer. The single crystal sample was mounted on its corners between the transducers to ensure nearly-free elastic boundary conditions for the sample.

We use the output of a custom-built lockin amplifier to excite one transducer at a fixed frequency and detect the quadrature response of the other transducer. We measure both in-phase (X) and out-of-phase (Y) components of the response. We achieve a full frequency sweep by stepping the drive frequency from about 100~kHz to 5~MHz. More details on the technique can be found in \cite{Ramshaw2015a, Shekhter2013a}.

\begin{figure}
	\centering
	\includegraphics[width=\linewidth]{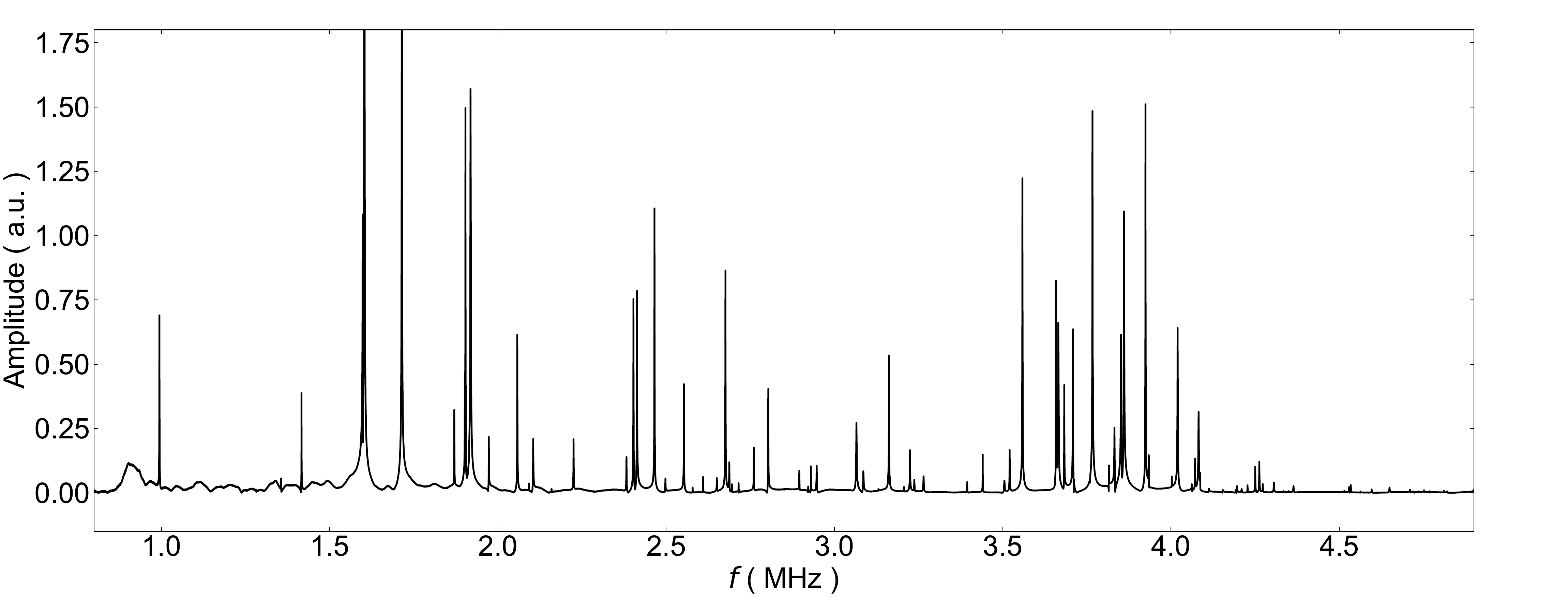}
	\caption{Amplitude of the response of the receiving transducer in the RUS measurement of \mnsn at room temperature. }
	\label{fig: raw scan}
\end{figure}

\autoref{fig: raw scan} shows the amplitude (X$^2+$Y$^2$) of an exemplary frequency sweep. We can identify mechanical resonances of the sample as frequencies at with maximum transmission between the drive and receive transducers occurs. From the position of these resonances we determine the full elastic tensor by inverse solving the elastic wave equation \cite{Ramshaw2015a,Shekhter2013a}. Lists of all experimental resonances included in the fit, alongside the calculated resonances and their differences, are shown in \autoref{table:mn3ge fit high T} for \mnge at 387~K and in \autoref{table:mn3sn fit high T} for \mnsn at 438~K. Data at room temperature for both compounds are shown in \autoref{table:RUS RT fit}.

Each resonance is a function of the density and dimensions of the sample, as well as all elastic moduli. We quantify the composition, $\alpha_{i,\mu}$, of each resonance $f_i$ by the logarithmic derivative with respect to the elastic moduli $c_\mu$
\begin{equation}
	\label{eq:logarithmic derivative}
	\alpha_{i,\mu} = \frac{\partial \left( \ln f_i^2 \right)}{\partial \left( \ln c_\mu \right)},
\end{equation}
with $\sum_\mu \alpha_{i,\mu} = 1$. These $\alpha$-coefficients are essentially geometric factors and depend only insignificantly on temperature. The temperature dependence of the resonance frequencies is therefore entirely determined by the temperature dependence of the elastic moduli and we can write
\begin{equation}
	\label{eq: frequency decomposition}
	\frac{2 \Delta f_i (T)}{f_i^0} = \sum_\mu \alpha_{i,\mu} \frac{\Delta c_\mu}{c_\mu^0}.
\end{equation}
$f_i^0$ and $c_\mu^0$ are the values of resonance frequencies and elastic moduli at a reference temperature, respectively. We compute these $\alpha$-coefficients by taking logarithmic derivatives of the calculated resonance frequencies at the elastic moduli returned by our fit. The $\alpha$-coefficients we received from our fits are given in \autoref{table:mn3ge fit high T} for \mnge\autoref{table:mn3sn fit high T} for \mnsn. With these values and \autoref{eq: frequency decomposition} we determined the temperature dependence of all elastic moduli. For this decomposition, we used the resonances displayed in bold font in \autoref{table:mn3ge fit high T} for \mnge and \autoref{table:mn3sn fit high T} for \mnsn. The reference temperatures were 387~K and 438~K for \mnge, \mnsn respectively. See \cite{Ramshaw2015a} and for more details on the used algorithm.

\begin{longtable}{|c|c|c|c|w{c}{1.75cm}|w{c}{1.75cm}|w{c}{1.75cm}|w{c}{1.75cm}|w{c}{1.75cm}|}
	\caption{Comparison between measured and calculated resonant ultrasound frequencies of \mnge at 387~K below 4~MHz. Also included are the $\alpha_i$ coefficients defined in the text. The resonances used to extract the temperature dependence of the elastic moduli are highlighted in bold font.\label{table:mn3ge fit high T}}\\
	\hline
	\multirow{2}{*}{Index}&\multirow{2}{*}{$f_{exp}$ (MHz)}&\multirow{2}{*}{$f_{calc}$ (MHz)}&\multirow{2}{*}{$\frac{f_{exp}-f_{calc}}{f_{calc}}$ (\%)}&\multicolumn{5}{c|}{$\alpha_i$}\\ 
																												\cline{5-9}
			  &      	&		&		&	$c_{11}$ & $c_{12}$ & $c_{13}$ & $c_{33}$ & $c_{44}$\\ 
	\hline
	\endfirsthead
	\multicolumn{9}{r}{Table 1 continued.}\\
	\hline
	\multirow{2}{*}{Index}&\multirow{2}{*}{$f_{exp}$ (MHz)}&\multirow{2}{*}{$f_{calc}$ (MHz)}&\multirow{2}{*}{$\frac{f_{exp}-f_{calc}}{f_{calc}}$ (\%)}&\multicolumn{5}{c|}{$\alpha_i$}\\ 
																												\cline{5-9}
	          &      	&		&		&	$c_{11}$ & $c_{12}$ & $c_{13}$ & $c_{33}$ & $c_{44}$\\ 
	\hline
	\endhead
	\hline
	\multicolumn{9}{r}{Table 1 continued on next page.}
	\endfoot
	\hline
	\endlastfoot
	1 &0.86329 &0.859066&0.492&0.032984&-0.009727&-0.000194&0.002945&0.973992\\
    \textbf{2} &\textbf{1.315918}&\textbf{1.312581}&\textbf{0.254}&\textbf{0.859018}&\textbf{-0.258943}&\textbf{-0.001859}&\textbf{0.013788}&\textbf{0.387997}\\
    3 &1.337735&1.336578&0.087&0.025095&-0.001121&-0.015063&0.578388&0.4127  \\
    \textbf{4 }&\textbf{1.359745}&\textbf{1.360394}&\textbf{0.048}&\textbf{0.017811}&\textbf{-0.003566}&\textbf{-4.6e-05 }&\textbf{0.032306}&\textbf{0.953495}\\
    \textbf{5 }&\textbf{1.380895}&\textbf{1.380585}&\textbf{0.022}&\textbf{0.02422 }&\textbf{-0.002075}&\textbf{-0.012132}&\textbf{0.552788}&\textbf{0.437198}\\
    \textbf{6 }&\textbf{1.451318}&\textbf{1.453035}&\textbf{0.118}&\textbf{0.015849}&\textbf{-0.003611}&\textbf{-0.000599}&\textbf{0.04641 }&\textbf{0.941951}\\
    \textbf{7 }&\textbf{1.627631}&\textbf{1.625203}&\textbf{0.149}&\textbf{1.2432  }&\textbf{-0.250281}&\textbf{-0.001504}&\textbf{0.002037}&\textbf{0.006547}\\
    \textbf{8 }&\textbf{1.652097}&\textbf{1.659973}&\textbf{0.474}&\textbf{1.409948}&\textbf{-0.416095}&\textbf{-0.002229}&\textbf{0.004276}&\textbf{0.004099}\\
    \textbf{9 }&\textbf{1.66751 }&\textbf{1.663834}&\textbf{0.221}&\textbf{1.076557}&\textbf{-0.293472}&\textbf{0.0009   }&\textbf{0.012452}&\textbf{0.203563}\\
    \textbf{10}&\textbf{1.706839}&\textbf{1.70703 }&\textbf{0.011}&\textbf{0.137226}&\textbf{-0.040441}&\textbf{-0.002095}&\textbf{0.121975}&\textbf{0.783335}\\
    \textbf{11}&\textbf{1.819579}&\textbf{1.815017}&\textbf{0.251}&\textbf{0.963598}&\textbf{-0.178073}&\textbf{-0.004892}&\textbf{0.002592}&\textbf{0.216775}\\
    \textbf{12}&\textbf{1.858302}&\textbf{1.858903}&\textbf{0.032}&\textbf{1.37357 }&\textbf{-0.400093}&\textbf{-0.007918}&\textbf{0.034419}&\textbf{2.3e-05 }\\
    \textbf{13}&\textbf{1.876659}&\textbf{1.872775}&\textbf{0.207}&\textbf{0.576531}&\textbf{-0.17169 }&\textbf{-0.005129}&\textbf{0.423329}&\textbf{0.176958}\\
    \textbf{14}&\textbf{1.905238}&\textbf{1.903118}&\textbf{0.111}&\textbf{0.884917}&\textbf{-0.232898}&\textbf{-0.029921}&\textbf{0.374152}&\textbf{0.00375 }\\
    15&1.914726&1.914616&0.006&1.247008&-0.250742&-0.005643&0.003007&0.006369\\
    \textbf{16}&\textbf{1.950258}&\textbf{1.950708}&\textbf{0.023}&\textbf{0.11054 }&\textbf{-0.014419}&\textbf{-0.042559}&\textbf{0.946302}&\textbf{0.000136}\\
    \textbf{17}&\textbf{2.043762}&\textbf{2.039117}&\textbf{0.228}&\textbf{1.023181}&\textbf{-0.191857}&\textbf{-0.004232}&\textbf{0.002602}&\textbf{0.170306}\\
    18&2.056432&2.05242 &0.195&0.542958&0.01066  &-0.00479 &0.03145 &0.419722\\
    19&2.073754&2.078511&0.229&0.616261&-0.176206&0.004173 &0.040856&0.514916\\
    20&2.154024&2.148777&0.244&1.203616&-0.214431&0.002949 &0.006192&0.001674\\
    21&2.259128&2.25193 &0.32 &0.660819&-0.121567&-0.011537&0.15332 &0.318965\\
    22&2.267891&2.264419&0.153&0.676047&-0.191125&-0.004776&0.047786&0.472068\\
    23&2.2821  &2.2756  &0.286&0.55479 &0.057397 &-0.026701&0.380458&0.034056\\
    24&2.28672 &2.280567&0.27 &0.999551&-0.195978&0.003161 &0.006039&0.187228\\
    25&2.331612&2.324629&0.3  &0.635782&-0.146923&-0.009937&0.297417&0.22366 \\
    26&2.353409&2.35653 &0.132&0.610414&-0.134801&-0.026279&0.189332&0.361334\\
    27&2.397702&2.396495&0.05 &0.207297&-0.005276&-0.024829&0.401571&0.421237\\
    28&2.405233&2.406206&0.04 &0.618556&-0.173377&-0.012234&0.106407&0.460647\\
    29&2.452348&2.451764&0.024&0.058538&0.000556 &-0.020601&0.46404 &0.497468\\
    30&2.481088&2.481642&0.022&0.077938&-0.011507&-0.002642&0.190769&0.745442\\
    31&2.495769&2.489697&0.244&0.231393&-0.043563&-0.000198&0.14338 &0.668988\\
    32&2.510943&2.520272&0.37 &0.869881&0.095786 &0.015274 &0.018786&0.000273\\
    33&2.573004&2.577864&0.189&0.116465&-0.032392&-0.004262&0.08384 &0.83635 \\
    34&2.587117&2.583748&0.13 &0.742318&-0.216434&-0.003558&0.20805 &0.269624\\
    35&2.645635&2.648888&0.123&1.372588&-0.387362&0.003314 &0.005892&0.005568\\
    36&2.660829&2.662691&0.07 &1.376685&-0.382699&0.000341 &0.002439&0.003233\\
    \textbf{37}&\textbf{2.707036}&\textbf{2.714622}&\textbf{0.279}&\textbf{0.743963}&\textbf{-0.138938}&\textbf{-0.005455}&\textbf{0.013762}&\textbf{0.386668}\\
    38&2.718903&2.721322&0.089&0.589388&0.068319 &-0.043898&0.169697&0.216494\\
    39&2.787169&2.78735 &0.007&0.765107&-0.211509&-0.001634&0.124491&0.323545\\
    40&2.819064&2.820408&0.048&0.265233&-0.049584&-0.016158&0.277128&0.523381\\
    \textbf{41}&\textbf{2.853757}&\textbf{2.853428}&\textbf{0.012}&\textbf{0.887737}&\textbf{-0.232789}&\textbf{-0.000429}&\textbf{0.102413}&\textbf{0.243068}\\
    \textbf{42}&\textbf{2.86775 }&\textbf{2.868243}&\textbf{0.017}&\textbf{0.91841 }&\textbf{-0.238958}&\textbf{0.00118  }&\textbf{0.099834}&\textbf{0.219534}\\
    \textbf{43}&\textbf{2.900467}&\textbf{2.899245}&\textbf{0.042}&\textbf{0.386185}&\textbf{-0.091994}&\textbf{0.000409 }&\textbf{0.208981}&\textbf{0.496419}\\
    44&3.004964&3.005587&0.021&0.359328&-0.083974&-0.007906&0.145206&0.587345\\
    45&3.027684&3.026964&0.024&1.310751&-0.315981&-0.004508&0.004083&0.005656\\
    46&3.055466&3.059585&0.135&0.422307&-0.094359&-0.004813&0.040347&0.636518\\
    47&3.075397&3.081272&0.191&0.31477 &-0.079211&-0.010581&0.29473 &0.480293\\
    48&3.12169 &3.11315 &0.274&0.543292&-0.134132&-0.003806&0.064057&0.530589\\
    49&3.153971&3.148271&0.181&1.204414&-0.291646&-0.005426&0.004638&0.08802 \\
    50&3.180459&3.184064&0.113&0.43201 &-0.100325&-0.00605 &0.059706&0.614658\\
    51&3.195999&3.191377&0.145&0.28304 &-0.026032&-0.014769&0.258002&0.499759\\
    52&3.236595&3.232265&0.134&0.720348&-0.181751&-0.009722&0.229515&0.241609\\
    53&3.302157&3.306776&0.14 &0.200287&-0.024516&-0.011027&0.214357&0.620899\\
    54&3.30783 &3.313241&0.163&0.252727&-0.019567&-0.001369&0.056688&0.711521\\
    55&3.311863&3.314773&0.088&0.409369&-0.119136&-0.005179&0.183026&0.53192 \\
    56&3.326765&3.32714 &0.011&0.474414&-0.069577&-0.005922&0.057524&0.543561\\
    57&3.338793&3.342154&0.101&0.365215&0.027917 &-0.0166  &0.142774&0.480694\\
    58&3.350336&3.365099&0.439&0.348317&0.039335 &-0.024843&0.234792&0.4024  \\
    59&3.406242&3.407761&0.045&0.590309&-0.150195&-0.00575 &0.167131&0.398504\\
    60&3.418442&3.419212&0.023&0.554145&-0.137717&-0.007238&0.178096&0.412715\\
    61&3.445046&3.448244&0.093&0.516247&-0.144521&0.002292 &0.04622 &0.579761\\
    62&3.453315&3.455402&0.06 &0.607016&-0.165121&-0.002184&0.155945&0.404345\\
    63&3.503158&3.499268&0.111&0.921779&-0.222359&-0.000231&0.014963&0.285849\\
    64&3.533018&3.53394 &0.026&1.32555 &-0.338575&-0.000523&0.002263&0.011285\\
    65&3.547987&3.544604&0.095&0.775429&-0.195812&-0.002863&0.104393&0.318853\\
    66&3.580287&3.571457&0.247&0.349965&-0.047269&-0.011251&0.197761&0.510793\\
    67&3.590199&3.593931&0.104&1.300358&-0.313532&0.00305  &0.003696&0.006427\\
    68&3.596618&3.594777&0.051&1.306984&-0.316448&-0.001609&0.00233 &0.008743\\
    69&3.648969&3.644795&0.115&0.370836&-0.054618&-0.006011&0.09036 &0.599432\\
    70&3.669138&3.664341&0.131&0.964061&-0.249704&-0.002499&0.028128&0.260014\\
    71&3.688207&3.698064&0.267&1.089497&-0.272837&-0.006796&0.10783 &0.082306\\
    72&3.698328&3.704204&0.159&1.387403&-0.394601&-0.00028 &0.001196&0.006282\\
    73&3.713795&3.709839&0.107&1.056882&-0.24787 &-0.001197&0.025127&0.167058\\
    74&3.73026 &3.730008&0.007&0.335607&-0.072146&-0.013116&0.214757&0.534897\\
    75&3.743308&3.741472&0.049&0.663521&-0.17438 &-0.009488&0.080292&0.440055\\
    76&3.750437&3.75735 &0.184&0.243461&-0.033184&7.6e-05  &0.128925&0.660721\\
    77&3.76102 &3.769173&0.216&0.570905&-0.160033&-0.004466&0.099212&0.494381\\
    78&3.763257&3.770313&0.187&0.343293&-0.068851&-0.006934&0.105501&0.626991\\
    79&3.811571&3.817756&0.162&0.661246&-0.135833&-0.014345&0.105957&0.382975\\
    80&3.860859&3.870459&0.248&0.750196&-0.19166 &-0.001334&0.09019 &0.352608\\
    81&3.869497&3.880479&0.283&1.378376&-0.383978&4.9e-05  &0.001478&0.004074\\
    82&3.898889&3.901128&0.057&0.406955&-0.0138  &-0.08308 &0.639288&0.050638\\
    83&3.942837&3.949442&0.167&0.01619 &0.000286 &0.001566 &0.980755&0.001204\\
    84&3.953787&3.953558&0.006&0.477156&-0.094303&-0.002564&0.116788&0.502923\\
\end{longtable}	

\begin{longtable}{|c|c|c|c|w{c}{1.75cm}|w{c}{1.75cm}|w{c}{1.75cm}|w{c}{1.75cm}|w{c}{1.75cm}|}
	\caption{Comparison between measured and calculated resonant ultrasound frequencies of \mnsn at 438~K below 4~MHz. Also included are the $\alpha_i$ coefficients defined in the text. The resonances used to extract the temperature dependence of the elastic moduli are highlighted in bold font.\label{table:mn3sn fit high T}}\\
	\hline
	\multirow{2}{*}{Index}&\multirow{2}{*}{$f_{exp}$ (MHz)}&\multirow{2}{*}{$f_{calc}$ (MHz)}&\multirow{2}{*}{$\frac{f_{exp}-f_{calc}}{f_{calc}}$ (\%)}&\multicolumn{5}{c|}{$\alpha_i$}\\ 
																												\cline{5-9}
			  &      	&		&		&	$c_{11}$ & $c_{12}$ & $c_{13}$ & $c_{33}$ & $c_{44}$\\ 
	\hline
	\endfirsthead
	\multicolumn{9}{r}{Table 2 continued.}\\
	\hline
	\multirow{2}{*}{Index}&\multirow{2}{*}{$f_{exp}$ (MHz)}&\multirow{2}{*}{$f_{calc}$ (MHz)}&\multirow{2}{*}{$\frac{f_{exp}-f_{calc}}{f_{calc}}$ (\%)}&\multicolumn{5}{c|}{$\alpha_i$}\\ 
																												\cline{5-9}
			  &      	&		&		&	$c_{11}$ & $c_{12}$ & $c_{13}$ & $c_{33}$ & $c_{44}$\\ 
	\hline
	\endhead
	\hline
	\multicolumn{9}{r}{Table 2 continued on next page.}\\
	\endfoot
	\hline
	\endlastfoot
	1 &0.956019&0.951489&0.476&0.023813&-0.004889&-0.000347&0.003037&0.978386\\
    \textbf{2 }&\textbf{1.302051}&\textbf{1.303615}&\textbf{0.12 }&\textbf{0.02655 }&\textbf{0.002418 }&\textbf{-0.035601}&\textbf{0.721671}&\textbf{0.284962}\\
    \textbf{3 }&\textbf{1.356757}&\textbf{1.356171}&\textbf{0.043}&\textbf{0.025468}&\textbf{0.001705 }&\textbf{-0.032502}&\textbf{0.706837}&\textbf{0.298492}\\
    4 &1.545377&1.547471&0.135&0.013534&-0.001769&-0.000944&0.057954&0.931226\\
    \textbf{5 }&\textbf{1.572799}&\textbf{1.559773}&\textbf{0.835}&\textbf{0.756634}&\textbf{-0.16096 }&\textbf{-0.00436 }&\textbf{0.02426 }&\textbf{0.384425}\\
    \textbf{6 }&\textbf{1.653398}&\textbf{1.652554}&\textbf{0.051}&\textbf{0.013347}&\textbf{-0.001924}&\textbf{-0.002878}&\textbf{0.087988}&\textbf{0.903467}\\
    7 &1.826925&1.825464&0.08 &0.067478&0.01191  &-0.083748&1.004346&1.4e-05 \\
    \textbf{8 }&\textbf{1.84124 }&\textbf{1.842193}&\textbf{0.052}&\textbf{0.287282}&\textbf{-0.05853 }&\textbf{-0.013344}&\textbf{0.2584  }&\textbf{0.526192}\\
    \textbf{9 }&\textbf{1.846365}&\textbf{1.850493}&\textbf{0.223}&\textbf{1.120941}&\textbf{-0.127257}&\textbf{-0.005996}&\textbf{0.003742}&\textbf{0.008571}\\
    \textbf{10}&\textbf{1.945893}&\textbf{1.947269}&\textbf{0.071}&\textbf{1.195594}&\textbf{-0.246653}&\textbf{-0.004122}&\textbf{0.007   }&\textbf{0.048182}\\
    11&1.951689&1.95092 &0.039&0.937129&-0.160907&-0.000914&0.023374&0.201318\\
    \textbf{12}&\textbf{2.004886}&\textbf{2.004361}&\textbf{0.026}&\textbf{0.43486 }&\textbf{-0.092516}&\textbf{-0.005243}&\textbf{0.273041}&\textbf{0.389857}\\
    13&2.045961&2.051807&0.285&0.886224&-0.086303&-0.01405 &0.007506&0.206623\\
    \textbf{14}&\textbf{2.064675}&\textbf{2.056601}&\textbf{0.393}&\textbf{0.824158}&\textbf{-0.132235}&\textbf{-0.043176}&\textbf{0.337083}&\textbf{0.01417 }\\
    \textbf{15}&\textbf{2.176206}&\textbf{2.183397}&\textbf{0.329}&\textbf{1.244708}&\textbf{-0.254064}&\textbf{0.0035   }&\textbf{0.005836}&\textbf{2e-05   }\\
    \textbf{16}&\textbf{2.198152}&\textbf{2.193   }&\textbf{0.235}&\textbf{1.105457}&\textbf{-0.121662}&\textbf{-0.015626}&\textbf{0.009977}&\textbf{0.021853}\\
    17&2.336228&2.326735&0.408&0.942624&-0.090457&-0.016915&0.011257&0.153491\\
    18&2.342467&2.331079&0.489&0.629228&0.019144 &-0.015867&0.05511 &0.312385\\
    19&2.375163&2.376649&0.063&0.592152&-0.120388&0.00801  &0.081065&0.439161\\
    20&2.396917&2.385549&0.477&0.63799 &0.045388 &-0.04756 &0.304204&0.05998 \\
    \textbf{21}&\textbf{2.448669}&\textbf{2.446057}&\textbf{0.107}&\textbf{1.079766}&\textbf{-0.101438}&\textbf{0.003177 }&\textbf{0.015105}&\textbf{0.003389}\\
    \textbf{22}&\textbf{2.477646}&\textbf{2.48024 }&\textbf{0.105}&\textbf{0.617914}&\textbf{-0.062612}&\textbf{-0.021414}&\textbf{0.152038}&\textbf{0.314075}\\
    \textbf{23}&\textbf{2.489611}&\textbf{2.483812}&\textbf{0.233}&\textbf{0.152701}&\textbf{-0.024537}&\textbf{-0.017044}&\textbf{0.362619}&\textbf{0.526262}\\
    24&2.553737&2.562227&0.331&0.665239&-0.079359&-0.064501&0.288148&0.190473\\
    25&2.584251&2.584635&0.015&0.06328 &0.004873 &-0.023037&0.342289&0.612596\\
    26&2.600278&2.59731 &0.114&0.662668&-0.053243&-0.027198&0.165498&0.252274\\
    27&2.605038&2.598638&0.246&0.899843&-0.110036&0.008528 &0.020661&0.181004\\
    28&2.611152&2.61477 &0.138&0.398005&-0.07819 &-0.013154&0.169501&0.523837\\
    29&2.626227&2.625597&0.024&0.146804&-0.024382&-0.010881&0.265314&0.623145\\
    \textbf{30}&\textbf{2.66705 }&\textbf{2.664576}&\textbf{0.093}&\textbf{0.256193}&\textbf{-0.034532}&\textbf{-0.005905}&\textbf{0.188479}&\textbf{0.595765}\\
    31&2.74096 &2.749941&0.327&0.888219&0.085302 &0.010094 &0.015804&0.00058 \\
    32&2.817082&2.804064&0.464&0.576538&-0.08574 &-0.017487&0.127791&0.398898\\
    33&2.820105&2.826817&0.237&0.113866&-0.020651&-0.008188&0.083303&0.83167 \\
    \textbf{34}&\textbf{2.85504 }&\textbf{2.862191}&\textbf{0.25 }&\textbf{0.604615}&\textbf{0.061504 }&\textbf{-0.099408}&\textbf{0.32672 }&\textbf{0.106569}\\
    \textbf{35}&\textbf{2.887556}&\textbf{2.881883}&\textbf{0.197}&\textbf{0.626516}&\textbf{-0.124236}&\textbf{-0.008912}&\textbf{0.171757}&\textbf{0.334875}\\
    36&3.053394&3.05536 &0.064&0.696949&-0.071694&-0.011713&0.013764&0.372694\\
    37&3.075785&3.077928&0.07 &0.325551&-0.042092&-0.016867&0.194129&0.539279\\
    38&3.083823&3.091586&0.251&1.230089&-0.236048&0.001915 &0.001843&0.002201\\
    39&3.097191&3.100364&0.102&1.152088&-0.210357&-0.007851&0.018753&0.047367\\
    40&3.124406&3.129561&0.165&0.35436 &-0.056171&-0.008243&0.297747&0.412307\\
    41&3.147232&3.142479&0.151&0.717732&-0.132002&-0.004214&0.111096&0.307389\\
    42&3.182915&3.188902&0.188&0.786225&-0.127707&-0.009681&0.147205&0.203957\\
    43&3.225322&3.226403&0.034&0.90845 &-0.151713&-0.005599&0.123719&0.125143\\
    44&3.301824&3.312192&0.313&0.354164&-0.057864&-0.022233&0.310435&0.415499\\
    45&3.33583 &3.343304&0.224&0.284697&-0.029478&-0.014191&0.136315&0.622657\\
    46&3.432684&3.434956&0.066&0.277038&-0.012263&-0.020699&0.260937&0.494987\\
    47&3.434088&3.444785&0.311&0.48636 &-0.080856&-0.004513&0.042928&0.556081\\
    48&3.47998 &3.484042&0.117&1.133368&-0.168501&-0.015017&0.0173  &0.03285 \\
    49&3.546648&3.556318&0.272&0.202206&-0.013989&-0.017091&0.240021&0.588853\\
    50&3.551306&3.559578&0.232&0.378032&0.038095 &-0.051217&0.308553&0.326536\\
    51&3.590234&3.590606&0.01 &0.51057 &-0.08022 &-0.010458&0.102737&0.477371\\
    52&3.600708&3.603821&0.086&0.409763&-0.062951&-0.012543&0.090234&0.575498\\
    53&3.616276&3.613859&0.067&1.072792&-0.158965&-0.015238&0.014565&0.086846\\
    54&3.643436&3.638197&0.144&0.628774&-0.106047&-0.015635&0.193127&0.29978 \\
    55&3.648954&3.648317&0.017&0.336665&-0.065239&-0.004221&0.163704&0.569091\\
    56&3.655188&3.653837&0.037&0.279852&0.020026 &-0.02265 &0.134945&0.587828\\
    57&3.702148&3.710773&0.232&0.316359&-0.012312&-0.02921 &0.313972&0.41119 \\
    58&3.728572&3.73323 &0.125&0.471791&-0.056456&-0.006404&0.070918&0.520151\\
    59&3.747068&3.746082&0.026&0.049625&-0.001996&0.012188 &0.910087&0.030096\\
    60&3.767859&3.770405&0.068&0.616607&-0.102735&-0.008799&0.177118&0.317808\\
    61&3.782426&3.781795&0.017&0.582054&-0.094426&-0.009466&0.183786&0.338052\\
    62&3.839776&3.84523 &0.142&0.479715&-0.086239&-0.014059&0.211026&0.409557\\
    63&3.876186&3.8759  &0.007&0.330193&-0.0189  &-0.016217&0.236648&0.468277\\
    64&3.891183&3.882874&0.214&0.240936&-0.010307&-0.066651&0.484947&0.351075\\
    65&3.935088&3.93423 &0.022&0.216267&-0.006227&-0.063648&0.6638  &0.189808\\
    66&3.962969&3.960474&0.063&0.385976&-0.05974 &-0.007909&0.171157&0.510516\\
    67&3.976983&3.979318&0.059&0.571261&-0.090972&-0.005854&0.125927&0.399638\\
    68&3.98941 &3.99035 &0.024&0.883511&-0.129195&-0.006181&0.008443&0.243421\\
\end{longtable}

\begin{longtable}{|c|c|c|c|c|c|c|}
	\caption{Room temperature RUS fit results for \mnge and \mnsn.\label{table:RUS RT fit}}\\
	\hline
	\multirow{4}{*}{Index} & \multicolumn{3}{c|}{\multirow{2}{*}{\mnge}} & \multicolumn{3}{c|}{\multirow{2}{*}{\mnsn}}\\
                           & \multicolumn{3}{c|}{} & \multicolumn{3}{c|}{}\\
						   \cline{2-7}
                          &\multirow{2}{*}{$f_{exp}$ (MHz)}&\multirow{2}{*}{$f_{calc}$ (MHz)}&\multirow{2}{*}{$\frac{f_{exp}-f_{calc}}{f_{calc}}$ (\%)}&\multirow{2}{*}{$f_{exp}$ (MHz)}&\multirow{2}{*}{$f_{calc}$ (MHz)}&\multirow{2}{*}{$\frac{f_{exp}-f_{calc}}{f_{calc}}$ (\%)}\\
    &&&&&&\\
	\hline
	\endfirsthead
	\multicolumn{7}{r}{Table 3 continued.}\\
	\hline
    \multirow{4}{*}{Index} & \multicolumn{3}{c|}{\multirow{2}{*}{\mnge}} & \multicolumn{3}{c|}{\multirow{2}{*}{\mnsn}}\\
                           & \multicolumn{3}{c|}{} & \multicolumn{3}{c|}{}\\
						   \cline{2-7}
                          &\multirow{2}{*}{$f_{exp}$ (MHz)}&\multirow{2}{*}{$f_{calc}$ (MHz)}&\multirow{2}{*}{$\frac{f_{exp}-f_{calc}}{f_{calc}}$ (\%)}&\multirow{2}{*}{$f_{exp}$ (MHz)}&\multirow{2}{*}{$f_{calc}$ (MHz)}&\multirow{2}{*}{$\frac{f_{exp}-f_{calc}}{f_{calc}}$ (\%)}\\
    &&&&&&\\
	\hline
	\endhead
	\hline
	\multicolumn{7}{r}{Table 3 continued on next page.}\\
	\endfoot
	\hline
	\endlastfoot
	1 &0.964686&0.956224&0.885&0.994195&0.988336&0.593\\
    2 &1.310108&1.317659&0.573&1.355897&1.360526&0.34 \\
    3 &1.467479&1.469985&0.17 &1.416341&1.415183&0.082\\
    4 &1.509547&1.509463&0.006&1.597958&1.580247&1.121\\
    5 &1.564017&1.560709&0.212&1.603094&1.60859 &0.342\\
    6 &1.578299&1.579394&0.069&1.714428&1.718097&0.214\\
    7 &1.584249&1.601805&1.096&1.870385&1.86784 &0.136\\
    8 &1.601411&1.613058&0.722&1.90144 &1.897641&0.2  \\
    9 &1.605816&1.618775&0.801&1.903465&1.908627&0.27 \\
    10&1.786309&1.800976&0.814&1.918534&1.94585 &1.404\\
    11&1.820358&1.811905&0.467&1.97281 &1.969372&0.175\\
    12&1.87303 &1.873366&0.018&2.057542&2.060389&0.138\\
    13&1.941317&1.931709&0.497&2.092095&2.086459&0.27 \\
    14&1.951124&1.953711&0.132&2.104788&2.090323&0.692\\
    15&2.033288&2.030126&0.156&2.159331&2.179086&0.907\\
    16&2.096093&2.083367&0.611&2.224686&2.214805&0.446\\
    17&2.105545&2.109193&0.173&2.381973&2.362535&0.823\\
    18&2.145873&2.123002&1.077&2.402662&2.401645&0.042\\
    19&2.162391&2.169167&0.312&2.413194&2.426592&0.552\\
    20&2.165406&2.176431&0.507&2.465173&2.46165 &0.143\\
    21&2.253102&2.264518&0.504&2.497626&2.47968 &0.724\\
    22&2.320023&2.307887&0.526&2.552512&2.540307&0.48 \\
    23&2.370485&2.350548&0.848&2.578463&2.572543&0.23 \\
    24&2.382754&2.382635&0.005&2.609505&2.621036&0.44 \\
    25&2.435864&2.418329&0.725&2.650515&2.639483&0.418\\
    26&2.462558&2.474235&0.472&2.675979&2.664001&0.45 \\
    27&2.473012&2.475588&0.104&2.687564&2.677365&0.381\\
    28&2.523838&2.549998&1.026&2.695202&2.690395&0.179\\
    29&2.554643&2.571023&0.637&2.715151&2.719726&0.168\\
    30&2.577582&2.590433&0.496&2.760339&2.759707&0.023\\
    31&2.61331 &2.622595&0.354&2.803317&2.832257&1.022\\
    32&2.654571&2.649171&0.204&2.89555 &2.868371&0.948\\
    33&2.719191&2.722728&0.13 &2.921852&2.929299&0.254\\
    34&2.751604&2.747202&0.16 &2.929966&2.938101&0.277\\
    35&2.780976&2.778539&0.088&2.947121&2.965036&0.604\\
    36&2.842243&2.839749&0.088&3.065267&3.090654&0.821\\
    37&2.860414&2.863792&0.118&3.085791&3.104826&0.613\\
    38&2.871657&2.870075&0.055&3.130633&3.121383&0.296\\
    39&2.911026&2.926243&0.52 &3.161843&3.176725&0.468\\
    40&2.931398&2.927636&0.128&3.206676&3.200352&0.198\\
    41&2.93455 &2.930862&0.126&3.224375&3.226009&0.051\\
    42&2.941563&2.94228 &0.024&3.23735 &3.240195&0.088\\
    43&3.044383&3.051906&0.246&3.264707&3.267632&0.09 \\
    44&3.10644 &3.095465&0.355&3.394271&3.414302&0.587\\
    45&3.12912 &3.133605&0.143&3.440456&3.454002&0.392\\
    46&3.193425&3.178438&0.472&3.504948&3.504163&0.022\\
    47&3.211411&3.211541&0.004&3.520484&3.534251&0.39 \\
    48&3.28658 &3.28379 &0.085&3.55844 &3.555479&0.083\\
    49&3.309553&3.297572&0.363&3.657986&3.642953&0.413\\
    50&3.315654&3.323133&0.225&3.665025&3.67687 &0.322\\
    51&3.357538&3.374973&0.517&3.682794&3.684646&0.05 \\
    52&3.38287 &3.378185&0.139&3.708454&3.693076&0.416\\
    53&3.441208&3.456592&0.445&3.710922&3.703852&0.191\\
    54&3.471899&3.4604  &0.332&3.759539&3.71424 &1.22 \\
    55&3.494208&3.488521&0.163&3.766532&3.754951&0.308\\
    56&3.498117&3.499481&0.039&3.767296&3.789343&0.582\\
    57&3.499642&3.512645&0.37 &3.815364&3.832352&0.443\\
    58&3.547463&3.531192&0.461&3.831993&3.839964&0.208\\
    59&3.572781&3.566142&0.186&3.851426&3.85365 &0.058\\
    60&3.584955&3.572166&0.358&3.860065&3.870464&0.269\\
    61&3.587674&3.598522&0.301&3.92407 &3.911587&0.319\\
    62&3.602309&3.60153 &0.022&3.933947&3.942623&0.22 \\
    63&3.607604&3.605513&0.058&4.002253&4.005757&0.087\\
    64&3.613076&3.60711 &0.165&4.019562&4.021003&0.036\\
    65&3.614359&3.614274&0.002&4.061745&4.046263&0.383\\
    66&3.618646&3.620746&0.058&4.071424&4.068287&0.077\\
    67&3.647651&3.642217&0.149&4.081862&4.075636&0.153\\
    68&3.670184&3.64649 &0.65 &4.086442&4.089984&0.087\\
    69&3.717184&3.701586&0.421&        &        &     \\
    70&3.72327 &3.723366&0.003&        &        &     \\
    71&3.787605&3.796285&0.229&        &        &     \\
    72&3.811398&3.806552&0.127&        &        &     \\
    73&3.8443  &3.822001&0.583&        &        &     \\
    74&3.869884&3.849273&0.535&        &        &     \\
    75&3.917229&3.915088&0.055&        &        &     \\
    76&3.945626&3.923404&0.566&        &        &     \\
    77&3.979077&3.976995&0.052&        &        &     \\
    78&3.999422&4.00085 &0.036&        &        &     \\
    79&4.01915 &4.009941&0.23 &        &        &     \\
    80&4.034952&4.023893&0.275&        &        &     \\
    81&4.045951&4.061753&0.389&        &        &     \\
    82&4.057585&4.065825&0.203&        &        &     \\
    83&4.064907&4.085995&0.516&        &        &     \\
    84&4.078004&4.09122 &0.323&        &        &     \\
\end{longtable}

\subsection{Samples Used in Measurements}
All samples used in our measurements were cut from one large \mnge and one \mnsn crystal. Final samples were polished to the shape of parallel prisms, with corners oriented along high symmetry directions. Dimensions of the samples are given below in the format $(a\times b \times c)$, where $a$ and $b$ are in-plane directions and $c$ is parallel to the c-axis.
For \mnge, we cut one $(915\times2575\times3080)~\mu$m piece for our pulse echo ultrasound measurements and one $(911\times1020\times1305)~\mu$m piece for our resonant ultrasound spectroscopy (RUS) measurements. This RUS sample was used for our fit at 387~K and to measure the temperature dependence of the elastic moduli. For the fit at room temperature, this sample was further polished to $(869\times1010\times1193)~\mu$m.
All RUS measurements on \mnsn were performed on a $(743\times836\times1.136)~\mu$m piece cut out of the original crystal.

\subsection{Poisson Ratios and Bulk Modulus}
\begin{figure}
	\centering
	\includegraphics[width=\linewidth]{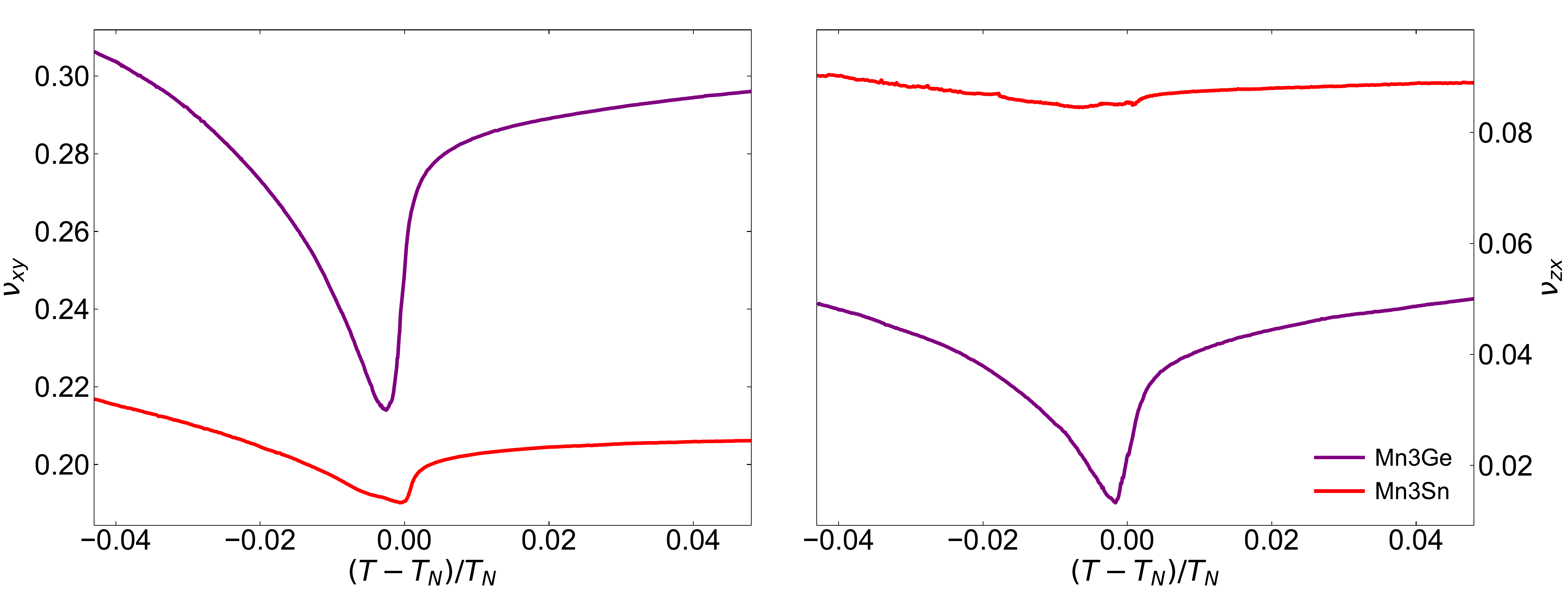}
	\caption{We show the Poisson ratios $\nu_{xy}$ (left panel) and $\nu_{zx}$ for \mnge (purple) and \mnsn (red) as a function of reduced temperature $(T-T_N)/T_N$. In a hexagonal crystal $\nu_{zx}=\nu_{zy}$.}
	\label{fig: poisson ratios}
\end{figure}

\begin{figure}
	\centering
	\includegraphics[width=.6\linewidth]{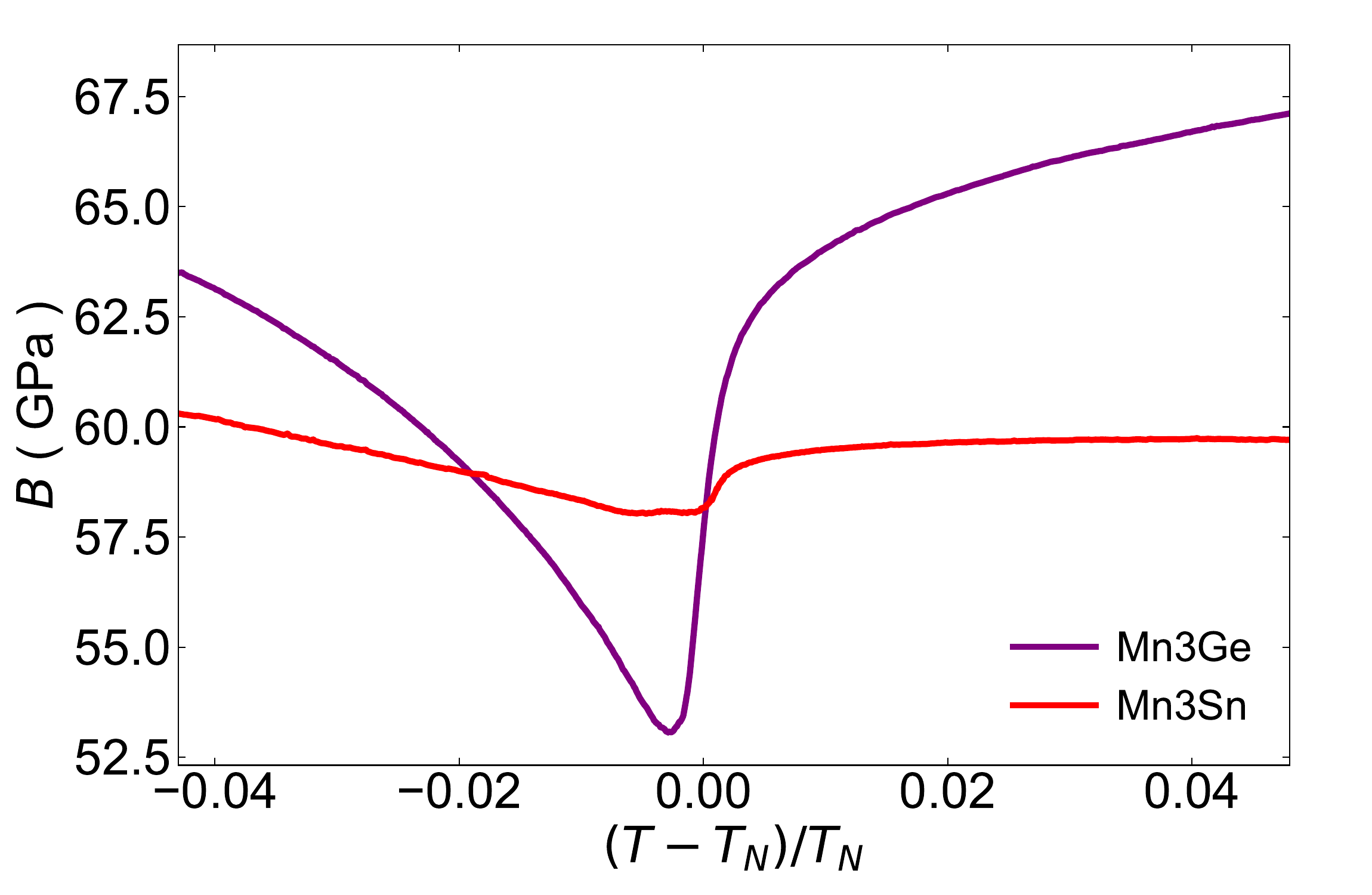}
	\caption{We show the bulk moduli for \mnge (purple) and \mnsn (red) as a function of reduced temperature $(T-T_N)/T_N$.}
	\label{fig: bulk moduli}
\end{figure}

Here we show the full temperature dependence of the Poisson ratios $\nu_{xy}$ and $\nu_{zx}$ (\autoref{fig: poisson ratios}), and bulk moduli (\autoref{fig: bulk moduli}) in \mnge and \mnsn.

\subsection{Landau Free Energy}
Elastic moduli are thermodynamic quantities defined as the second derivative of the free energy with respect to strain. The precursor fluctuations above $T_N$ in \mnx indicate that there substantial, non-mean-field corrections to the thermodynamics near the phase transition. However, defining a Landau free energy is still useful to illustrate the symmetry of the coupling terms and the expected behavior of the moduli ``not too close'' to the phase transition. 

 The free energy, $\mathscr{F}$, relevant to our measurements can be divided into an elastic part $f_{el}$, the free energy for the order parameter $f_{OP}$, a term considering the coupling between order parameter and strain $f_{coupling}$, a Zeeman-term $f_{Zeeman}$, and $f_{piezo}$---a term trilinear in order parameter, magnetic field, and $E_{2g}$ strain. The total free energy is then
\begin{equation}
	\mathscr{F} = f_{el} + f_{OP} + f_{coupling} + f_{Zeeman} + f_{piezo}.
	\label{SI:general_free_energy}
\end{equation}
These parts will be discussed separately in the following subsections.

\subsubsection{Elastic Free Energy and Poisson's ratio}
The elastic tensor only has five independent elements in $D_{6h}$. In Voigt notation, it reads
\begin{equation*}
	c = 
	\begin{pmatrix}
	c_{11} & c_{12} & c_{13} &  0     &     0  &  0  \\
	c_{12} & c_{11} & c_{13} &  0     &     0  &  0  \\
	c_{13} & c_{13} & c_{33} &  0     &     0  &  0  \\
	0      & 0      & 0      & c_{44} &     0  &  0  \\
	0      & 0      & 0      &  0     & c_{44} &  0  \\
	0      & 0      & 0      &  0     &     0  &  \frac{ c_{11}-c_{12} }{2}  
	\end{pmatrix}.
\end{equation*}
With a strain vector defined as $\varepsilon = \left\{ \varepsilon_{xx}, \varepsilon_{yy}, \varepsilon_{zz}, 2 \varepsilon_{yz}, 2 \varepsilon_{xz}, 2 \varepsilon_{xy} \right\}$, the elastic free energy in $D_{6h}$ is
\begin{align}
	f_{el} &= \frac{1}{2} \varepsilon_{i} c_{ij} \varepsilon_{j} \\
	       &= \frac{1}{2} \left( 2c_{12} \left(-\varepsilon_{xy}^2 + \varepsilon_{xx} \varepsilon_{yy} \right) + 
		   c_{11} \left(\varepsilon_{xx}^2 + 2 \varepsilon_{xy}^2 + \varepsilon_{yy}^2 \right) \right. + 4 c_{44} \left( \varepsilon_{xz}^2 + \varepsilon_{yz}^2 \right) + \left. 2c_{13} \left( \varepsilon_{xx} + \varepsilon_{yy} \right) \varepsilon_{zz} + c_{33} \varepsilon_{zz}^2 \right) \\
		   \
		   &= \frac{1}{2} \left( \frac{c_{11}-c_{12}}{2} \left( \varepsilon_{xx}+\varepsilon_{yy} \right)^2 + c_{33} \varepsilon_{zz}^2 \right. + 2 c_{13}\varepsilon_{zz}\left( \varepsilon_{xx}+\varepsilon_{yy} \right) + 4 c_{44} \left( \varepsilon_{xz}^2 + \varepsilon_{yz}^2 \right) \left. + \frac{c_{11}-c_{12}}{2} \left( \left( \varepsilon_{xx}-\varepsilon_{yy} \right)^2 + 4 \varepsilon_{xy}^2 \right) \right)\\
		   &= \frac{1}{2} \left( c_{A1g,1} \varepsilon_{A1g,1}^2 + c_{A1g,2} \varepsilon_{A1g,2}^2 \right. \left.+ 2 c_{A1g,3} \varepsilon_{A1g,1} \varepsilon_{A1g,2} + c_{E1g} \left| \boldsymbol{\varepsilon}_{E1g} \right|^2 + c_{E2g} \left| \boldsymbol{\varepsilon}_{E2g} \right|^2 \right).
	\label{SI:elastic_free_energy}
\end{align}
Here, the irreducible strains $\varepsilon_\Gamma$ are defined as
\begin{align}
	\varepsilon_{A1g,1} &= \varepsilon_{xx} + \varepsilon_{yy}, \\
	\varepsilon_{A1g,2} &= \varepsilon_{zz}, \\
	\boldsymbol{\varepsilon}_{E1g}   &= \left\{ 2\varepsilon_{xz},\, 2\varepsilon_{yz} \right\}, \\
	\boldsymbol{\varepsilon}_{E2g}   &= \left\{ \varepsilon_{xx}-\varepsilon_{yy},\, 2\varepsilon_{xy} \right\}. 
\end{align}
These strains are linear combinations of elements of the strain tensor $\varepsilon_{ij}$ and are the physically relevant quantities as they transform as irreducible representations $\Gamma$ with respect to the $D_{6h}$ point group. The elastic moduli $c_\Gamma$ corresponding to these strains are
\begin{align}
	c_{A1g,1} &= \frac{c_{11} + c_{12}}{2}, \\
	c_{A1g,2} &= c_{33}, \\
	c_{A1g,3} &= c_{13}, \\
	c_{E1g}   &= c_{44}, \\
	c_{E2g}   &= \frac{c_{11} - c_{12}}{2}.
\end{align}
In a hexagonal crystal, in--plane and out--of--plane Poisson's ratios are given by
\begin{align}
	\nu_{xy} &= \frac{c_{13}^2 - c_{12} c_{33}}{c_{13}^2 - c_{11} c_{33}}, \\
	\nu_{zx} = \nu_{zy} &= \frac{\left(c_{11} - c_{12}\right) c_{13}}{-c_{13}^2 + c_{11} c_{33}}.
	\label{SI:poisson_ratio}
\end{align}
The bulk modulus is defined in terms of elastic moduli as
\begin{equation}
    \label{eq: bulk modulus}
    B = \frac{\frac{c_{11}+c_{12}}{2} c_{33} - c_{13}^2}{\frac{c_{11} + c_{12}}{2} + c_{33} - 2 c_{13}}.
\end{equation}

\subsubsection{Order Parameter Free Energy}

The order parameter that forms in both \mnsn and \mnge, at the high-temperature $T_N$ studied here, is of the $E_{1g}$ representation \cite{Chen2020,Soh2020a}. It is therefore a two-component order parameter that can be written as $\boldsymbol{\eta} = \left\{ \eta_x, \eta_y \right\}$. Up to fourth order in $\boldsymbol{\eta}$, the Landau free energy is
\begin{equation}
	f_{OP} = \alpha \left( T - T_N \right) \left| \boldsymbol{\eta} \right|^2 + \beta_1 \left| \boldsymbol{\eta} \right|^4 + \beta_2 \left( \eta_x^2 - \eta_y^2 \right) + \beta_3 \eta_x^2 \eta_y^2.
\end{equation}
Hexagonal crystal symmetry requires $b_3 = 4 b_2$, which simplifies the free energy to
\begin{equation}
	f_{OP} = \alpha \left(T-T_N\right) \eta^2 + \beta \eta^4,
\end{equation}
with $b = b_1 + b_2$, and where we have parametrized the order parameter as $\boldsymbol{\eta} = \eta \left\{ \cos \left(\phi_\eta\right), \sin \left(\phi_\eta\right) \right\}$.. Note that this free energy is isotropic---sixth-order is the lowest order at which anisotropy appears. 



\subsubsection{Coupling of Order Parameter and Strain in the Free Energy}

The order parameter has to appear in even powers because it breaks time reversal symmetry. The allowed couplings between strain and order parameter are
\begin{align}
	f_{coupling} &= \sum_{i=1}^3 \gamma_{A1g,i} \varepsilon_{A1g,i} \left|\boldsymbol{\eta}^2 \right| + \gamma_{E1g} \left|\boldsymbol{\varepsilon}_{E1g}^2 \right| \left|\boldsymbol{\eta}^2 \right|+ \gamma_{E2g} \left( \varepsilon_{E2g,x} \left( \eta_x^2-\eta_y^2 \right) + 2\varepsilon_{E2g,y} \eta_x\eta_y  \right)\\
	&= \sum_{i=1}^3 \gamma_{A1g,i} \varepsilon_{A1g,i} \eta^2 + \gamma_{E1g} \left|\boldsymbol{\varepsilon}_{E1g}^2 \right| \eta^2 + \gamma_{E2g} \eta^2 \varepsilon_{E2g} \cos \left( 2 \left( \phi_\varepsilon - \phi_\eta \right)\right),
\end{align}
where we have used the parametrization of the order parameter given above, as well as $\boldsymbol{\varepsilon}_{E2g} = \varepsilon_{E2g} \left\{ \cos \left( 2 \phi_\varepsilon \right), \sin \left( 2 \phi_\varepsilon \right) \right\}$.

\subsubsection{Zeeman Energy}
At zero applied strain, the total magnetization $\boldsymbol{M}$ is proportional to the order parameter:
\begin{equation}
    \label{eq:magnetic moment}
    \boldsymbol{M} = \delta \boldsymbol{\eta},
\end{equation}
where $\delta$ is a coefficient. The Zeeman term in the free energy in the presence of an in-plane magnetic field, $\boldsymbol{H}=h \left\{ \cos\left( \phi_h\right), \sin \left(\phi_h\right)\right\}$, is given by
\begin{align}
    \label{eq:Zeeman energy}
    f_{Zeeman} &= - \delta \boldsymbol{\eta} \boldsymbol{H}\\
    &=-\delta \eta h \cos \left( \phi_\eta - \phi_h \right).
\end{align}

\subsubsection{Piezomagnetic Term}
Both the order parameter and an in-plane magnetic field break time-reversal symmetry and transform as the $E_{1g}$ representation. Thus, a term in the free energy which is trilinear in order parameter, magnetic field, and $A_{1g}$ or $E_{2g}$ strain is allowed by symmetry. Forming all $A_{1g}$ products of order parameter, field, and strain, we find
\begin{equation}
    \label{eq:piezo free energy}
    f_{piezo} = \sum_{i=1}^{2} \lambda_{A1g,i} \varepsilon_{A1g,i}\eta h \cos \left( \phi_\eta - \phi_h \right) + \lambda \varepsilon_{E2g} \eta h \cos \left( \phi_\eta + \phi_h - 2 \phi_\varepsilon  \right),
\end{equation}
using the same polar coordinates as above.

\subsubsection{Full Free Energy}
Combining all the term discussed above, the full free energy is given by
\begin{align}
	\label{eq:full free energy}
	\mathscr{F} &= \frac{1}{2} \left( c_{A1g,1} \varepsilon_{A1g,1}^2 + c_{A1g,2} \varepsilon_{A1g,2}^2 \right. \left.+ 2 c_{A1g,3} \varepsilon_{A1g,1} \varepsilon_{A1g,2} + c_{E1g} \left| \boldsymbol{\varepsilon}_{E1g} \right|^2 + c_{E2g} \left| \boldsymbol{\varepsilon}_{E2g} \right|^2 \right)\\
	\notag
	&\quad + \alpha \left(T-T_N\right) \eta^2 + \beta \eta^4\\
	\notag
	&\quad + \sum_{i=1}^3 \gamma_{A1g,i} \varepsilon_{A1g,i} \eta^2 + \gamma_{E1g} \left|\boldsymbol{\varepsilon}_{E1g}^2 \right| \eta^2 + \gamma_{E2g} \eta^2 \varepsilon_{E2g} \cos \left( 2 \left( \phi_\varepsilon - \phi_\eta \right)\right)\\
	\notag
	&\quad -\delta \eta h \cos \left( \phi_\eta - \phi_h \right)\\
	\notag
	&\quad - \sum_{i=1}^{2} \lambda_{A1g,i} \varepsilon_{A1g,i}\eta h \cos \left( \phi_\eta - \phi_h \right) - \lambda \varepsilon_{E2g} \eta h \cos \left( \phi_\eta + \phi_h - 2 \phi_\varepsilon  \right).
\end{align}

\subsection{Elastic Constants at the Phase Transition}
To estimate the behaviour of the elastic moduli through the phase transition, we consider the free energy at zero field, and constant angles $\phi_h$, $\phi_\eta$, and $\phi_\varepsilon$, which are the constraints of our ultrasound measurements. In this case, the free energy given by \autoref{eq:full free energy} simplifies to
\begin{align}
	\label{eq:full free energy without angles}
	\mathscr{F} &= \frac{1}{2} \left( c_{A1g,1} \varepsilon_{A1g,1}^2 + c_{A1g,2} \varepsilon_{A1g,2}^2 \right. \left.+ 2 c_{A1g,3} \varepsilon_{A1g,1} \varepsilon_{A1g,2} + c_{E1g} \left| \boldsymbol{\varepsilon}_{E1g} \right|^2 + c_{E2g} \left| \boldsymbol{\varepsilon}_{E2g} \right|^2 \right)\\
	\notag
	&\quad + \alpha \left(T-T_N\right) \eta^2 + \beta \eta^4 + \sum_{i=1}^3 \gamma_{A1g,i} \varepsilon_{A1g,i} \eta^2 + \gamma_{E1g} \left|\boldsymbol{\varepsilon}_{E1g}^2 \right| \eta^2 + \gamma_{E2g} \eta^2 \varepsilon_{E2g}\\
	\notag
	&\quad - \sum_{i=1}^{2} \lambda_{A1g,i} \varepsilon_{A1g,i}\eta h  - \lambda \varepsilon_{E2g} \eta h ,
\end{align}
where the cosine terms from \autoref{eq:full free energy} are absorbed into the expansion coefficients. We find the equilibrium order parameter, $\eta_{eq}$, through $\left. \left( d \mathscr{F}/ d \eta \right) \right|_{\eta_{eq}} = 0$, and the elastic moduli $c_{\Gamma}$ are defined through $\left. \left(\partial^2 \mathscr{F} / \partial \varepsilon_\Gamma^2 \right) \right|_{\eta_{eq}}$.

For $\varepsilon_{A1g}$ and $\varepsilon_{E2g}$ strains, which couple linearly to the square of the order parameter as $\varepsilon_\Gamma \eta^2$, we find a step discontinuity in the temperature dependence of their respective elastic moduli at the phase transition
\begin{equation}
	\label{eq: elastic constants jump at Tc}
	\Delta c_\Gamma = \left(c_\Gamma \left( T>T_N \right) - c_\Gamma \left(T<T_N\right) \right)_{T \rightarrow T_N} = \frac{2 \gamma_\Gamma^2}{\beta}.
\end{equation}
However, for $c_{E1g}$ whose corresponding strain couples to the order parameter as $\left|\varepsilon_{E1g}\right|^2 \eta^2$ to lowest order, this mean-field analysis yields
\begin{equation}
	\label{eq: elastic constants kink at Tc}
	\Delta c_{E1g} = \left[c_{E1g} \left( T>T_N \right) - c_{E1g} \left(T<T_N\right) \right]_{T \rightarrow T_N} = 0.
\end{equation}
We therefore expect to see only a change in slope in the temperature dependence of $c_{E1g}$ at $T_N$.